\documentclass[journal]{IEEEtran} 
\usepackage{mathtools, amsthm, thmtools, amssymb} 
\usepackage{newtxmath}

\usepackage{cite}
\usepackage{orcidlink} 
\usepackage[right, mathlines]{lineno}  

\usepackage{subcaption}
\usepackage{graphicx}
\usepackage[all, poly, knot]{xy}

\theoremstyle{plain}
\newtheorem*{AGW}{AGW's Criterion}
\newtheorem{construction}{Construction} 
\newtheorem{theorem}{Theorem}  
\newtheorem{lemma}{Lemma}
\newtheorem{corollary}{Corollary}

\theoremstyle{definition}
\newtheorem{definition}{Definition}
\newtheorem{example}{Example}

\theoremstyle{remark}
\newtheorem{remark}{Remark}

\usepackage{xcolor}
\hypersetup{
    colorlinks = true,      
    citecolor = violet,       
    linkcolor = violet,      
    urlcolor = purple,       
    pdfborderstyle = {/S/U/W 0}  
}

\usepackage[shortlabels]{enumitem}  

\usepackage{doi}
\usepackage[nameinlink,capitalize]{cleveref} 
\crefname{theorem}{Theorem}{Theorems}
\crefname{lemma}{Lemma}{Lemmas}
\crefname{corollary}{Corollary}{Corollaries}
\crefname{construction}{Construction}{Constructions}
\crefformat{equation}{#2\textup{(#1)}#3}
\crefmultiformat{equation}{(#2#1#3)}{ and~(#2#1#3)}{, (#2#1#3)}{ and~(#2#1#3)}
\crefformat{figure}{Fig.\:#2#1#3} 

\usepackage{xspace}  

\newcommand{\ifa}{if and only if\xspace}
\newcommand\qtq[1]{\quad\text{#1}\quad\xspace}

\newcommand{\ord}{\mathrm{ord}}

\newcommand{\onetoone}{$1$-to-$1$\xspace}
\newcommand{\twotoone}{$2$-to-$1$\xspace}
\newcommand{\thrtoone}{$3$-to-$1$\xspace}

\newcommand\mtoone{$m$-to-$1$\xspace}
\newcommand\ntoone{$n$-to-$1$\xspace}

\newcommand\mfq{$m$-to-$1$ on~$\mathbb{F}_{q}$\xspace}
\newcommand\mfqstar{$m$-to-$1$ on~$\mathbb{F}_{q}^{*}$\xspace}

\newcommand\mfqtwostar{$m$-to-$1$ on $\mathbb{F}_{q^2}^{*}$\xspace}

\newcommand\mfield[2]{$#1$-to-$1$ on~$#2$\xspace}
\newcommand\mset[2]{$#1$-to-$1$ on~$#2$\xspace}

\newcommand{\tr}{\mathrm{Tr}}
\newcommand{\trtnt}{\mathrm{Tr}_{2^n/2}}

\newcommand{\trqnq}{\mathrm{Tr}_{q^n/q}}

\newcommand{\n}{\mathbb{N}}
\newcommand{\z}{\mathbb{Z}}

\newcommand{\f}{\mathbb{F}}

\newcommand{\ftwon}{\mathbb{F}_{2^n}}

\newcommand{\ftwonstar}{\mathbb{F}_{2^n}^{*}}

\newcommand{\fq}{\mathbb{F}_{q}}
\newcommand{\fqx}{\mathbb{F}_{q}[x]}
\newcommand{\fqstar}{\mathbb{F}_{q}^{*}}

\newcommand{\fqtwo}{\mathbb{F}_{q^2}}
\newcommand{\fqtwox}{\mathbb{F}_{q^2}[x]}
\newcommand{\fqtwostar}{\mathbb{F}_{q^2}^{*}}

\newcommand{\fqn}{\mathbb{F}_{q^n}}

\newcommand{\fqnstar}{\mathbb{F}_{q^n}^{*}}

\begin{document}

\title{On Many-to-One Mappings Over Finite Fields}
\author{
Yanbin Zheng\textsuperscript{\orcidlink{0009-0009-1177-4515}}, 
Yanjin Ding, Meiying Zhang, Pingzhi Yuan\textsuperscript{\orcidlink{https://orcid.org/0000-0001-6398-0614}}, and 
Qiang Wang\textsuperscript{\orcidlink{0000-0001-5426-2776}}  \\[9pt]
In Memory of Professor Dingyi Pei (1941--2023) \vspace{-18pt}%
\thanks{
 This work was supported by 
 the National Key R\&D Program of China (No.\ 2024YFA1013000), 
 the National Natural Science Foundation of China 
    (Nos.\ 12571103, 12461103),
 and NSERC of Canada (RGPIN-2023-04673).
}%
\thanks{Y. Zheng, Y. Ding and M. Zhang are with the School of Mathematical Sciences, Qufu Normal University, Qufu, China.  
Y. Zheng is also with the Institute of Information Engineering, Chinese Academy of Sciences, Beijing, China (e-mail: zheng@qfnu.edu.cn). Y. Ding is also with the Hubei Key Laboratory of Applied Mathematics, Faculty of Mathematics and Statistics, Hubei University, Wuhan, China. }%
\thanks{P. Yuan is with the School of Mathematical Sciences, South China Normal University, Guangzhou, China (e-mail: yuanpz@scnu.edu.cn).}%
\thanks{Q. Wang is with the School of Mathematics and Statistics, Carleton University, Ottawa, Canada (e-mail: wang@math.carleton.ca).}%
\thanks{This paper has been published in IEEE Transactions on Information Theory. Please refer to this paper as: 
\href{https://doi.org/10.1109/TIT.2026.3657349}{Y. Zheng, Y. Ding, M. Zhang, P. Yuan, Q. Wang, On many-to-one mappings over finite fields, IEEE
Trans. Inf. Theory, vol.~72, no.~5, pp.~2734--2752, May, 2026.}}
}



\maketitle

\begin{abstract}
\boldmath
We introduce the definition of $m$-to-$1$ mappings between two finite sets, which unifies and generalizes the definitions of $2$-to-$1$ and $n$-to-$1$ mappings in recent literature. We also characterize these $m$-to-$1$ mappings in terms of the generalized local criterion and thus provide three generic constructions of $m$-to-$1$ mappings, which unify and generalize the previous known constructions. Using these constructions, the problem whether $x^r h(x^s)$ is $m$-to-$1$ on the multiplicative group $\mathbb{F}_{q}^{*}$ is converted into that whether an associated polynomial $x^{r_1} h(x)^{s_1}$ is $m_2$-to-$1$ on the order~$\ell$ subgroup~$U_{\ell}$ of $\mathbb{F}_{q}^{*}$, where $m_2 = m / (r, s)$ and $\ell = (q-1) / s$. Furthermore, the $m_2$-to-$1$ property of $x^{r_1} h(x)^{s_1}$ on $U_{\ell}$ is studied in detail in four different cases. In addition, a recursive construction of $m$-to-$1$ mappings from $m$-to-$1$ mappings is proposed. 
\end{abstract}


\section{Introduction}

\IEEEPARstart{O}{ne-to-one} mappings from a finite field $\fq$ 
to itself (i.e., permutations of $\fq$) 
have been extensively studied; see for example 
\cite{Hou15, WangIndex19, Wang25survey, Zheng5deg} 
and the references therein. Two-to-one mappings over~$\fq$ 
are closely connected to linear codes 
\cite{Ding153265, Ding162288} and cryptographic 
functions such as APN functions, planar functions, and bent functions
\cite{Dob99Welch, GoharP08Planar, WengZ12PlanarDO, BentFunctions, Gohar11, Charpin14}.  
This motivates the further theoretical study of two-to-one mappings;
see for example \cite{2to1-MesQ19, 2to1-LiSQ21, Kolsch25}. 
In order to unify and generalize previous results in the 
literature, this paper provides a systematic study of \mtoone 
(a generalization of two-to-one) mappings over~$\fq$ including 
their characterization, properties, and construction methods.

\subsection{The progress of many-to-one mappings}

\begin{figure}[!t]
    \centering
    \includegraphics[width=0.45\textwidth]{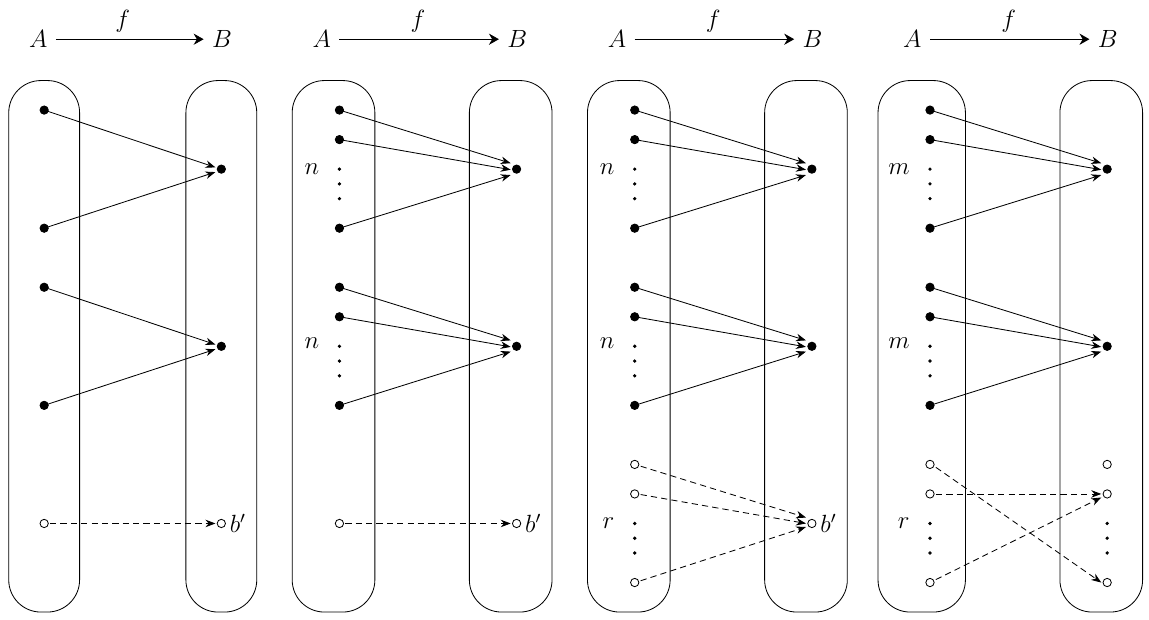} \\
        {\footnotesize  \hspace{0.1cm} 
        \twotoone \hspace{0.8cm}	
        \ntoone in~\cite{GAO21mto1}	  \hspace{0.3cm}
        \ntoone in~\cite{nto1-NiuLQL23} \hspace{0.5cm}
        our \mtoone }
    \caption{Schematic diagrams of many-to-one mappings}
    \label{fig:defs_mto1}
\end{figure}

Assume $A$ and $B$ are finite sets 
and $f$ is a mapping from $A$ to~$B$. 
For any $b \in B$, let $\# f^{-1}(b)$ denote 
the number of preimages of~$b$ under $f$. 
The mapping~$f$ is called \twotoone if 
$\# f^{-1}(b) \in \{0, 2\}$ for each $b \in B$, 
except for at most a single $b' \in B$ 
such that $\# f^{-1}(b') = 1$;
see the first column of \cref{fig:defs_mto1}.

\begin{figure*}[!t]
\centering
\begin{subfigure}{0.3\linewidth}
	\centering
    \[
    \xymatrix{
      A \ar[rrr]^{f ~~ \text{\twotoone}}
        _{f|_{\lambda^{-1}(s)} ~~\text{\color{blue}\twotoone}}
        \ar[d]_{\lambda}  
      &   &   &  A  \ar[d]^{\bar{\lambda}} \\
      S \ar[rrr]^{\bar{f} ~~ \text{\onetoone}}           
      &   &   &  \bar{S}}
    \]\\[-9pt]  
	\caption{\twotoone in~\cite{2to1-MesQ19}}
	\label{fig:21Sihem}
\end{subfigure}
\begin{subfigure}{0.3\linewidth}
	\centering
    \[
    \xymatrix{
      A \ar[rrr]^{f ~~ \text{\ntoone}}  
        _{f|_{\lambda^{-1}(s)} ~~\text{\color{blue}\ntoone}}
        \ar[d]_{\lambda}   
      &  &   &  A  \ar[d]^{\bar{\lambda}} \\
      S \ar[rrr]^{\bar{f} ~~ \text{\onetoone}}           
      &   &   &  \bar{S}}
    \]\\[-9pt] 
	\caption{\ntoone in~\cite{GAO21mto1, Qin2469}}
	\label{fig:n1Gao}
\end{subfigure}
\begin{subfigure}{0.3\linewidth}
	\centering
    \[
    \xymatrix{
      A \ar[rrr]^{f ~~ \text{\mtoone}}  
        _{f|_{\lambda^{-1}(s)} ~~\text{\color{red}\mtoone}}
        \ar[d]_{\lambda}   
      &  &   &  \bar{A}  \ar[d]^{\bar{\lambda}} \\
      S \ar[rrr]^{\bar{f} ~~ \text{\onetoone}}           
      &   &   &  \bar{S}}
    \]\\[-9pt] 
	\caption{Our \cref{constr1}}
	\label{fig:OurCons1}
\end{subfigure}
\\
\begin{subfigure}{0.3\linewidth}
	\centering
    \[
    \xymatrix{
      A \ar[rrr]^{f ~~ \text{\twotoone}} 
        _{f|_{\lambda^{-1}(s)} ~~\text{\onetoone}}
        \ar[d]_{\lambda}  
      &   &   &  \bar{A}  \ar[d]^{\bar{\lambda}} \\
      S \ar[rrr]^{\bar{f} ~~ {\text{\color{blue}\twotoone}}}          
      &   &   &  \bar{S}}
    \]\\[-9pt]  
	\caption{\twotoone in~\cite{2to1-YuanZW21}}
	\label{fig:21Yuan}
\end{subfigure}
\begin{subfigure}{0.3\linewidth}
	\centering
    \[
    \xymatrix{
      A \ar[rrr]^{f ~~ \text{\ntoone}}  
        _{f|_{\lambda^{-1}(s)} ~~\text{\onetoone}}
        \ar[d]_{\lambda}  
      &  &   &   A  \ar[d]^{\bar{\lambda}} \\
      S \ar[rrr]^{\bar{f} ~~ \text{\color{blue}\ntoone}}            
      &   &   &  \bar{S}}
    \]\\[-9pt]  
	\caption{\ntoone in~\cite{nto1-NiuLQL23}}
	\label{fig:n1Niu}
\end{subfigure}
\begin{subfigure}{0.3\linewidth}
	\centering
    \[
    \xymatrix{
      A \ar[rrr]^{f ~~ \text{\mtoone}}  
        _{f|_{\lambda^{-1}(s)} ~~ {\color{red}m_1\text{-to-}1} }
        \ar[d]_{\lambda}   
      &   &   &  \bar{A}  \ar[d]^{\bar{\lambda}} \\
      S \ar[rrr]^{\bar{f} ~~ {\color{red} m/m_1\text{-to-}1}}            
      &   &   &  \bar{S}}
    \]\\[-9pt]  
	\caption{Our \cref{constr2}}
	\label{fig:OurCons2}
\end{subfigure}
    \caption{Constructions of many-to-one mappings}
    \label{fig:mto1Constr}
\end{figure*}

The study of \twotoone mappings is partly motivated 
by the design of cryptographic functions, 
in particular APN and planar functions; see for example 
\cite{Dob99Welch, GoharP08Planar, WengZ12PlanarDO,
Charpin09, Gohar11, Charpin14}. 
In 2019, Mesnager and Qu~\cite{2to1-MesQ19} 
provided a systematic study of 
\twotoone mappings over finite fields 
and presented several constructions of \twotoone mappings 
by using an AGW-like criterion (see \cref{fig:21Sihem}), 
permutation polynomials, linear translators, and APN functions.
They also reviewed several classical types of \twotoone 
polynomial mappings, including Dickson polynomials, 
Muller-Cohen-Matthews polynomials, among others.
Moreover, all \twotoone polynomials of degree $\le 4$ 
over any finite field were determined in~\cite{2to1-MesQ19}. 
In 2021, Li et~al.\ \cite{2to1-LiSQ21} proved that
there is no \twotoone polynomials of degree~5 
over $\ftwon$ when $n > 9$, 
and found some classes of \twotoone  
trinomials and quadrinomials over $\ftwon$.
In the same year, Yuan et~al.\ \cite{2to1-YuanZW21}
presented another new AGW-like criterion 
(see \cref{fig:21Yuan}) for \twotoone mappings.
Using this criterion, they obtained new 
constructions of \twotoone mappings 
and eight classes of \twotoone mappings of the form 
$(x^{2^k} + x + \delta)^s + c x$ over $\ftwon$. 
In 2023, Mesnager et~al.\ \cite{MesYZ231315}
proposed several classes of \twotoone mappings of the form 
$x^r + x^s + x^t + x^3 + x^2 + x$,
$\sum_{i=1}^{2} (x^{2^k} + x + \delta)^{s_i} + c x$,
or $h(x) \circ (x^{2^e} + x)$ over $\ftwon$,
where $(e, n) = 1$ and $h$ is \onetoone on the 
image set of $x^{2^e} + x$. 
Very recently, K\"{o}lsch and Kyureghyan 
\cite{Kolsch25} observed that on $\ftwon$
the classification of \twotoone binomials is 
equivalent to the classification of o-monomials, 
which is a well-studied and elusive problem 
in finite geometry. They also provided some 
connections between \twotoone maps and hyperovals 
in non-Desarguesian planes. 
As applications, the \twotoone mappings over~$\ftwon$  
are used to construct cryptographic functions  
\cite{2to1-MesQ19, QuZL25JoC, ZhangLiQu26}
and linear codes \cite{KLiHQ214263, MesQC23719, MesQCY233285}.

In 2021, the concept of \twotoone mappings was 
generalized to \ntoone mappings~\cite{GAO21mto1}. 
More specifically, when $\# A \equiv 0, 1 \pmod{n}$,
$f$ is called an \ntoone mapping if 
$\# f^{-1}(b) \in \{0, n\}$ for each $b \in B$,
except for at most a single $b' \in B$ 
such that $\# f^{-1}(b') = 1$; 
see the second column of \cref{fig:defs_mto1}.
Later, an extended definition of \ntoone mappings 
was introduced in~\cite{BartGT22194} (on finite field~$A$) 
and independently in~\cite{nto1-NiuLQL23} (on finite set~$A$), 
which allows $\# A \bmod{n} \in \{0, 1, \ldots, n-1\}$. 
Specifically, $f$ is called an \ntoone mapping if 
$\# f^{-1}(b) \in \{0, n\}$ for each $b \in B$,
except for at most a single $b' \in B$ 
such that $\# f^{-1}(b') = r$, where $r = \#A \bmod n$; 
see the third column of \cref{fig:defs_mto1}. 
In this case,~$f$ maps the remaining~$r$ 
elements in $A$ to the same image~$b'$ if $r \ne 0$. 
Under this definition, a new method to obtain \mbox{\ntoone} 
mappings based on Galois groups of rational functions 
was proposed, and two explicit classes of 
\twotoone and \thrtoone rational functions over 
finite fields were given in~\cite{BartGT22194}. 
Very recently, the main result of~\cite{BartGT22194} was 
refined and generalized in~\cite{DingZ24}.
Under this definition, all \thrtoone polynomials 
of degree $\le 4$ over finite fields 
were determined in~\cite{nto1-NiuLQL23}.
Moreover, an AGW-like criterion (see \cref{fig:n1Niu})
for characterizing \ntoone mappings 
was presented in~\cite{nto1-NiuLQL23}, 
and this criterion was 
applied to polynomials of the forms
$h(\psi(x))\phi(x) + g(\psi(x))$, 
$L_1(x) + L_2(x) g(L_3(x))$, $x^r h(x^s)$ 
and $g(x^{q^k} - x + \delta) + c x$ over finite fields. 
Later, several classes of \ntoone trinomials of the form 
$x^r h(x^{q-1})$ over $\fqtwo$ were given in~\cite{Qin2469}.
Recently, a relationship between \mbox{\thrtoone} mappings and 
APN functions was given in \cite{KolschKK23, BudaghyanIK2335}. 
In particular, K{\"{o}}lsch et~al.\ \cite{KolschKK23} showed 
that APN functions on $\ftwon$ with the minimal image size 
are very close to being \mbox{\thrtoone}. Budaghyan et~al.\ 
\cite{BudaghyanIK2335} provided a list of 
all known quadratic \mbox{\thrtoone} APN functions 
on $\ftwon$ for $n \leq 12$ up to EA-equivalence.

All the definitions of \ntoone mappings in
\cite{GAO21mto1, BartGT22194, nto1-NiuLQL23} require that
$\# f(R) = 1$, where $f(R)$ denotes the image of the set~$R$
of the remaining~$r$ elements in $A$ if $r \ne 0$.
In this paper, we will introduce a more general concept,
\mtoone mappings, which allow that
$\# f(R) \in \{1, 2, \ldots, r\}$; 
see the last column of \cref{fig:defs_mto1}.

\subsection{The constructions of many-to-one mappings}
In this subsection, we will focus on the constructions 
of many-to-one mappings, which was motivated 
by the following well known AGW criterion 
\cite{AGW,TuxanidyWang14} for \onetoone mappings.

\begin{AGW} 
Let $A$, $\bar{A}$, $S$, $\bar{S}$ be finite sets 
with $\# A = \# \bar{A}$ and $\#S=\#\bar{S}$, 
and let $f \colon A\rightarrow \bar{A}$,
$\bar{f} \colon S\rightarrow\bar{S}$, $\lambda \colon A\rightarrow S$,
and $\bar{\lambda} \colon \bar{A}\rightarrow \bar{S}$ be maps 
such that $\bar{\lambda}\circ f= \bar{f}\circ\lambda$. 
If both~$\lambda$ and $\bar{\lambda}$ are surjective, 
then~$f$ is \onetoone from~$A$ to~$\bar{A}$ \ifa
$\bar{f}$ is \onetoone from~$S$ to~$\bar{S}$ 
and~$f$ is \onetoone on $\lambda^{-1}(s)$ for each $s \in S$.
\end{AGW}

The AGW criterion allows us to construct \onetoone maps on~$A$ 
from \onetoone maps between two smaller sets~$S$ and~$\bar{S}$,
which can be illustrated by the following commutative diagram:
\[
\xymatrix{
  A \ar[rrr]^{f ~~ \text{\onetoone}}
    _{f|_{\lambda^{-1}(s)} ~~\text{\onetoone}}
    \ar[d]_{\lambda}  
  &   &   &  \bar{A}  \ar[d]^{\bar{\lambda}} \\
  S \ar[rrr]^{\bar{f} ~~ \text{\onetoone}}           
  &   &   &  \bar{S}.
}
\]

In recent years, the AGW criterion has been generalized 
to construct \twotoone and \ntoone mappings in \cite{2to1-MesQ19, GAO21mto1, nto1-NiuLQL23, 2to1-YuanZW21, Qin2469}. 
All these constructions have the same assumption: 
$A$, $\bar{A}$, $S$, and $\bar{S}$ are finite sets,
and $f \colon A \rightarrow A$ (or $\bar{A}$),
$\bar{f} \colon S \rightarrow \bar{S}$, 
$\lambda \colon A \rightarrow S$, and 
$\bar{\lambda} \colon A~(\text{or}~\bar{A}) \rightarrow \bar{S}$ 
are mappings such that 
$\bar{\lambda}\circ f= \bar{f}\circ\lambda$. 
We now review these constructions.

\begin{itemize}
\item Proposition~6 in~\cite{2to1-MesQ19} states that, 
    if $\#S = \#\bar{S}$, 
    $\bar{f}$ is \onetoone from $S$ to $\bar{S}$, 
    $f|_{\lambda^{-1}(s)}$ is \twotoone for any $s \in S$, 
    and there is at most one $s \in S$ such that 
    $\#\lambda^{-1}(s)$ is odd, then $f$ is \mset{2}{A};
    see \cref{fig:21Sihem}.
\item Proposition~1 in~\cite{GAO21mto1} states that,  
    if $\# A \equiv 0, 1 \pmod{n}$, $\#S = \#\bar{S}$, 
    $\bar{f}$ is \onetoone from $S$ to $\bar{S}$, 
    $f|_{\lambda^{-1}(s)}$ is \ntoone for any $s \in S$, 
    and there is at most one $s \in S$ such~that 
    $\#\lambda^{-1}(s) \equiv 1 \pmod n$, 
    then $f$ is \mset{n}{A}; see \cref{fig:n1Gao}.
\item Theorem~3.1 in~\cite{Qin2469} states that, 
    if $\#S = \#\bar{S}$, 
    $\lambda$ and $\bar{\lambda}$ are surjective,
    $n \mid \# \lambda^{-1}(s)$, and $f|_{\lambda^{-1}(s)}$ 
    is \ntoone for any $s \in S$, 
    then $f$ is \mset{n}{A} \ifa $\bar{f}$ is \mset{1}{S}.
    However, there is a small error in their statement and  a 
    counter-example is the following: let $A = \f_{9}^*$, 
    $S = \bar{S} = \{\pm 1\}$, $f = \bar{f} = x^{6}$, 
    and $\lambda = \bar{\lambda} = x^4$. 
    It is easy to very that $\# \lambda^{-1}(s) = 4$ 
    and $f|_{\lambda^{-1}(s)}$ is \twotoone for any $s \in S$.
    But $f$ is \mset{2}{A} and $\bar{f}$ is also \mset{2}{S}. 
    The reason why their statement fails is that the condition
    $n \mid \# \lambda^{-1}(s)$ is too weak. However, the result 
    can be fixed by strengthening this condition to
    $\# \lambda^{-1}(s) = n \,\# \bar{\lambda}^{-1}(\bar{f}(s))$;
    see our \cref{mto1:m1=m}. 
\item Proposition~4.2 in {\cite{2to1-YuanZW21}} states that, 
    if $f$, $\bar{f}$, $\lambda$, $\bar{\lambda}$ are surjective,
    $f$ is \onetoone from $\lambda^{-1}(s)$ to 
    $\bar{\lambda}^{-1}(\bar{f}(s))$ for any $s \in S$,
    $\# S$ is even, and $\bar{f}$ is \twotoone from~$S$ 
    to~$\bar{S}$, then $f$ is \mset{2}{A}; see \cref{fig:21Yuan}.
    It should be noted that the above conditions 
    imply that $\# A$ is even;
    see the proof of our \cref{mto1_m2|S}. 
\item Theorem~4.3 in~\cite{nto1-NiuLQL23}  
    assumes that $\lambda$ and  $\bar{\lambda}$ are surjective,
    $\# A \equiv \#S  \pmod{n}$, $\#S = \#\bar{S}$,   
    and~$f$ is \onetoone from $\lambda^{-1}(s)$ to 
    $\bar{\lambda}^{-1}(\bar{f}(s))$ for any $s \in S$.
    When $n \mid \#S$, $f$ is \mset{n}{A} \ifa 
    $\bar{f}$ is \ntoone on $S$; see \cref{fig:n1Niu}. 
    When $n \nmid \#S$, they did not give necessary and 
    sufficient condition for $f$ to be \mset{n}{A}.
\end{itemize}

\subsection{The contribution of the paper}

The definition of \mtoone mappings between two finite sets 
is introduced in \cref{sec_def},
which unifies and generalizes the definitions 
of $2$-to-$1$ and $n$-to-$1$ mappings in the literature. 
\cref{sec_constr} presents a generalized local criterion,
which describes a necessary and sufficient condition of 
\mtoone mappings on finite sets. 
Then three constructions of \mtoone mappings are proposed 
by employing the generalized local criterion.
The first construction reduces the problem 
whether~$f$ is \mtoone on a finite set~$A$ 
to a relatively simpler problem 
whether~$f$ is \mtoone on some subsets of~$A$. 
The second construction converts the problem 
whether~$f$ is \mtoone on~$A$ into another problem 
whether an associated mapping $\bar{f}$ is 
$m / m_1$-to-$1$ on a finite set~$S$. 
These two constructions can be illustrated by 
\cref{fig:OurCons1,fig:OurCons2} respectively,
and they unify and generalize all the constructions 
of \twotoone mappings and \ntoone mappings in 
\cite{2to1-MesQ19, GAO21mto1, nto1-NiuLQL23, 2to1-YuanZW21, Qin2469}.  
In the case that $(A, \ast)$ is a finite group, 
the third construction reduces the problem whether 
$f \ast u$ is \mtoone on~$A$ to that whether~$f$ 
is \mtoone on~$A$, where $u$ is a mapping from~$A$ to~$A$
such that $\bar{\lambda}( u(a) )= c$ 
for any $a \in A$ and a fixed $c \in A$.

In \cref{sec_xrhxs}, applying the second construction,
the problem whether $f(x) \coloneqq x^r h(x^s)$ 
is \mtoone on the multiplicative group $\fqstar$ 
is converted into another problem whether 
$g(x) \coloneqq x^{r_1} h(x)^{s_1}$ 
is $m_2$-to-$1$ on the multiplicative 
subgroup~$U_{\ell}$; see \cref{MainThm} for details. 
In the following sections, the $m_2$-to-$1$ property of~$g$ on 
$U_{\ell}$ is explicitly discussed in four different cases: 
(1) $m = 2, 3$ or $\ell = 2, 3$;
(2) $g$ behaves like a monomial on $U_{\ell}$;
(3) $g$ behaves like a rational function on $U_{\ell}$;
(4) we repeatedly use the second construction and 
convert the problem that~$g$ is $m_2$-to-$1$ on~$U_{\ell}$ 
into that another associated mapping $\bar{g}$ 
is $m_3$-to-$1$ on a finite set~$\lambda(U_{\ell})$. 
These results and the connections with earlier results 
in the literature are presented from \cref{sec_small_ml} 
to \cref{sec_further_reduction}. 
Finally, we present two methods for constructing 
\mtoone mappings from  permutations or \mtoone mappings. 
This provides a recursive construction of \mtoone mappings. 

\subsection{Notation}
The letter $\z$ will denote the set of all integers,
$\n$ the set of all positive integers,
$\#S$ the cardinality of a finite set $S$,
and $\varnothing$ the empty set containing no elements. 
The greatest common divisor of two integers 
$a$ and $b$ is written as $(a, b)$. 
Let $a \bmod m$ denote the smallest non-negative 
remainder obtained when~$a$ is divided by~$m$. 
For a prime power $q$, 
let $\fq$ be the finite field~with~$q$ elements,
$\fqstar = \fq \setminus \{0\}$, 
and $\fqx$ the ring of polynomials over~$\fq$.
Let $U_{\ell}$ denote the cyclic group of 
all $\ell$-th roots of unity over $\fq$, i.e.,
$U_{\ell} = \{\alpha \in \fqstar : \alpha^\ell = 1\}$. 
The \emph{trace function from $\mathbb{F}_{q^n}$ to 
$\mathbb{F}_{q}$} is defined by
$\trqnq(x) = \sum_{i=0}^{n-1} x^{q^i}$.

\section{The definition of \texorpdfstring{\mtoone}{m-to-1} mappings}
\label{sec_def}

\begin{definition}\label{defn:mto1}
  Let $A$ be a finite set and $m \in \z$ with 
  $1 \le m \le \# A$. Write $\# A = k m + r$, 
  where $k$, $r \in \z$ with $0 \le r < m$. 
  Let~$f$ be a mapping from~$A$ to another finite set~$B$. 
  Then~$f$ is called \mtoone on~$A$ if 
  there are~$k$ distinct elements in~$B$ such that 
  each element has exactly~$m$ preimages in~$A$ under~$f$.
  The remaining~$r$ elements in~$A$ are called the
  \textit{exceptional elements} of~$f$ on~$A$,
  and the set of these~$r$ exceptional elements 
  is called the \textit{exceptional set} 
  of~$f$ on~$A$ and denoted by $E_{f}(A)$. 
  In particular, $E_{f}(A) = \varnothing$ 
  \ifa $r = 0$, i.e., $m \mid \# A$.
\end{definition}

In the case either $r = 0$ or $r \ne 0$ 
and $\# f(E_{f}(A)) = 1$, 
\cref{defn:mto1} is the same as the definition of \twotoone
or \mbox{\ntoone} mentioned in the previous section. 
In other cases, it is a generalization of 
the definitions in the literature. 
Throughout this paper, 
we use \cref{defn:mto1} in all of our results. 
Moreover, it should be noted that
$f$ is \onetoone on~$A$ means that
$f$ is \onetoone from~$A$ to $f(A)$,
where $f(A)$ may not equal $A$.
If~$f$ is \mbox{$m$-to-$1$} on~$A$, then any $b \in f(A)$ 
has at most~$m$ preimages in~$A$ under~$f$.

\begin{definition}
A polynomial $f(x) \in \fqx$ is called \mfq 
if the associated mapping 
$f \colon c \mapsto f(c)$ from $\fq$ to itself is \mfq.
\end{definition}

\begin{example}
Let $f(x) = x^3 + x$. Then $f$ maps $0,1,2,3,4$ 
to $0,2,0,0,3$ in~$\f_5$, respectively. 
Thus,  $f$ is \mfield{3}{\f_5} and the 
exceptional set $E_{f}(\f_5) = \{1, 4\}$.
\end{example}

\begin{example}\label{xnFq}
The monomial $x^n$ with $n \in \n$ is \mfield{(n, q-1)}{\fqstar}
and 
$E_{x^n}(\fqstar)= \varnothing$.
\end{example}

This example can be generalized to the following form.

\begin{example}\label{EndKer}
Let $f$ be an endomorphism of a finite group~$G$ and
$\ker(f) = \{ x \in G : f(x) = e\}$,
where $e$ is the identity element in~$G$. 
It is easy to verify that 
$\{ x \in G : f(x) = f(a)\} = a \ker(f)$ 
for any $a \in G$. Hence, $f$ is \mset{m}{G} and
$E_f(G) = \varnothing$, where $m = \# \ker(f)$.
\end{example}

We next compute the number of all \mtoone mappings on~$\fq$.

\begin{theorem}
Let $q = km + r$, where $1 \le m \le q$ and $0 \le r < m$. 
Let $N_{m}(\fq)$ be the number of all \mtoone mappings 
from $\fq$ to itself. Then 
\[
  N_{m}(\fq) = \frac{(q!)^2 \, (q-k)^r}{k! \, r! \, (m!)^k (q-k)!}.
\] 
\end{theorem}
\begin{proof}
For any \mtoone mapping $f$ on $\fq$ with  $q = km + r$, 
we get $\# E_{f}(\fq) = r$ and $\# f(\fq \setminus E_{f}(\fq)) = k$.
Then there are $\binom{q}{k}$ choices for $f(\fq \setminus E_{f}(\fq))$. 
For the first element in $f(\fq \setminus E_{f}(\fq))$, 
its preimage has $\binom{q}{m}$ choices. For the second element, 
its preimage has $\binom{q-m}{m}$ choices. 
Similar results hold until the last element has $\binom{m+r}{m}$ choices. 
Moreover, the image of each element in $E_{f}(\fq)$ has $q-k$ choices. Hence,  
\begin{align*}
  N_{m}(\fq) 
  & = \binom{q}{k} \binom{q}{m} \binom{q-m}{m} \cdots 
      \binom{m+r}{m} (q-k)^r  \\
  & = \binom{q}{k} \frac{q! \, (q-k)^r}{r! \, (m!)^k}.
  \qedhere
\end{align*}
\end{proof}

\section{Generalized local criterion and three constructions for \texorpdfstring{\mtoone}{m-to-1} mappings}\label{sec_constr}

Very recently, the local criterion 
for a mapping to be a permutation of~$A$ 
was provided by Yuan~\cite{Yuan24Local}, 
which is equivalent to the AGW criterion.
We now present a generalization of it 
for a mapping to be $m$-to-$1$ on~$A$.

\begin{lemma}[Generalized local criterion]\label{LocalLem}
Let $A$, $B$, and $C$ be finite sets. 
Let $f \colon A \to B$, $\psi \colon B \to C$, and $\varphi \colon A \to C$ 
be mappings such that $\varphi = \psi \circ f$, 
i.e., the following diagram is commutative:
\[
\xymatrix{   
    A  \ar@{->}[dr]_{\varphi} \ar[rr]^{f}   
    & &  B \ar@{->}[dl]^{\psi} \\
    & C &   
}
\] 
For any $c \in \varphi(A)$, let 
$\varphi^{-1}(c) = \{ a \in A : \varphi(a) = c \}$. 
Then for $1 \le m \le \#A$, 
$f$ is $m$-to-$1$ on~$A$ \ifa $f$ is \mfield{m}{\varphi^{-1}(c)} 
for any $\# \varphi^{-1}(c) \ge m$ and 
\begin{equation}\label{sumE=r}
    \sum_{\# \varphi^{-1}(c) \ge m} \# E_{f}(\varphi^{-1}(c)) 
  + \sum_{\# \varphi^{-1}(c) < m} \# \varphi^{-1}(c) = \#A \bmod{m},
\end{equation}
where $c$ runs through $\varphi(A)$, 
$E_{f}(\varphi^{-1}(c))$ is the exceptional set 
of~$f$ being \mfield{m}{\varphi^{-1}(c)}. 
\end{lemma}
\begin{proof}
Assume that $\varphi(A) = 
\{ c_1, c_2, \ldots, c_n \}$. Then 
\[
A = \varphi^{-1}(c_1) \uplus \varphi^{-1}(c_2) 
    \uplus\cdots\uplus \varphi^{-1}(c_n),
\]
where $\uplus$ denotes the union of disjoint sets. Thus,  
\[ 
    f(A) = f(\varphi^{-1}(c_1)) 
        \cup f(\varphi^{-1}(c_2)) 
        \cup\cdots\cup f(\varphi^{-1}(c_n)).
\]
By $\varphi = \psi \circ f$,  we have 
\[
\psi(f(\varphi^{-1}(c_i))) 
= \varphi(\varphi^{-1}(c_i)) = c_i,
\]
and hence
\[
  f(\varphi^{-1}(c_i)) \subseteq \psi^{-1}(c_i).
\]
If $c_i \ne c_j$, then 
$\psi^{-1}(c_i) \cap \psi^{-1}(c_j) = \varnothing$, and thus 
\[
f(\varphi^{-1}(c_i)) \cap f(\varphi^{-1}(c_j)) = \varnothing.
\]
Therefore, 
\begin{equation}\label{eq:fA=fc1+fcn}
    f(A) = f(\varphi^{-1}(c_1)) \uplus f(\varphi^{-1}(c_2)) \uplus\cdots\uplus f(\varphi^{-1}(c_n)).
\end{equation}

Let $\#A = km + r$, where $0 \le r < m$.
($\Leftarrow$) 
Assume $f$ is \mbox{$m$-to-$1$} on $\varphi^{-1}(c_i)$ 
for any $\# \varphi^{-1}(c_i) \ge m$ and \cref{sumE=r} holds. 
Then there are~$(\# A - r)/m = k$ distinct elements in~$f(A)$ 
such that each element has exactly~$m$ preimages 
in~$A$ under~$f$. Hence $f$ is $m$-to-$1$ on~$A$.
($\Rightarrow$) 
Assume $f$ is $m$-to-$1$ on~$A$. Then there are at most~$m$ 
preimages in $\varphi^{-1}(c_i)$ for any element in 
$f(\varphi^{-1}(c_i))$ and $\# E_{f}(A) = r < m$, 
where $E_{f}(A)$ is the exceptional set 
of $f$ being $m$-to-$1$ on~$A$.
If $\# \varphi^{-1}(c_i) \ge m$, 
let $\# \varphi^{-1}(c_i) = k_i m + r_i$  
with $k_i \ge 1$ and $0 \le r_i < m$, 
and let $k'_i$ be the number of $b \in f(\varphi^{-1}(c_i))$ 
which has exactly~$m$ preimages in $\varphi^{-1}(c_i)$. 
If $k'_i < k_i$, then 
\begin{align*}
\# E_{f}(\varphi^{-1}(c_i)) 
& = \# \varphi^{-1}(c_i) - k'_i m \\
& = (k_i - k'_i) m + r_i \ge m,
\end{align*}
contrary to $\# E_{f}(A) < m$. Thus $k'_i = k_i$, 
i.e., $f$ is \mfield{m}{\varphi^{-1}(c_i)} if 
$\# \varphi^{-1}(c_i) \ge m$. 
If $\# \varphi^{-1}(c_i) < m$, 
then $\varphi^{-1}(c_i) \subseteq E_{f}(A)$ by~\cref{eq:fA=fc1+fcn}. Hence,
\begin{equation}\label{EfA=Efphi}
\Big(\underset{\# \varphi^{-1}(c) \ge m}{\uplus} E_{f}(\varphi^{-1}(c))\Big) 
  \uplus \Big(\underset{\# \varphi^{-1}(c) < m}{\uplus}  \varphi^{-1}(c)\Big)
  = E_{f}(A)
\end{equation}
and thus \cref{sumE=r} holds. 
\end{proof}

The generalized local criterion converts the problem 
whether $f$ is \mtoone on~$A$ to another problem
 whether $f$ is \mtoone on some subsets $\varphi^{-1}(c)$ of~$A$. 
The identity \cref{sumE=r} establishes a numerical
relationship between the exceptional sets 
$E_{f}(A)$ and $E_{f}(\varphi^{-1}(c))$.
In fact, this relationship can also be expressed as  
\cref{EfA=Efphi} when $f$ is $m$-to-$1$ on~$A$.
We next use this criterion to deduce 
three constructions of \mtoone mappings.

\subsection{The first construction}

\begin{construction}\label{constr1}
Let $A$, $\bar{A}$, $S$, $\bar{S}$ be finite sets 
and $f \colon A \rightarrow \bar{A}$, 
$\bar{f} \colon S \rightarrow \bar{S}$, 
$\lambda \colon A \rightarrow S$, 
$\bar{\lambda} \colon \bar{A} \rightarrow \bar{S}$ 
be mappings such that 
$\bar{\lambda} \circ f = \bar{f} \circ \lambda$,
i.e., the following diagram is commutative:
\[
\xymatrix{
  A \ar[rr]^{f}\ar[d]_{\lambda}  &   &  
  \bar{A}  \ar[d]^{\bar{\lambda}} \\
  S \ar[rr]^{\bar{f}}            &   &  \bar{S}.}
\]
For any $s \in \lambda(A)$, let 
$\lambda^{-1}(s) = \{ a \in A : \lambda(a) = s \}$. 
Suppose~$\bar{f}$ is \mfield{1}{S}. 
Then for $1 \le m \le \#A$, $f$ is $m$-to-$1$ on~$A$ \ifa 
$f$ is \mfield{m}{\lambda^{-1}(s)} 
for any $\# \lambda^{-1}(s) \ge m$ and 
\[
  \sum_{\# \lambda^{-1}(s) \ge m} \# E_{f}(\lambda^{-1}(s)) 
  +\sum_{\# \lambda^{-1}(s) < m} \# \lambda^{-1}(s) = \#A \bmod{m},
\]
where $s$ runs through $\lambda(A)$, 
$E_{f}(\lambda^{-1}(s))$ is the exceptional set of~$f$ 
being \mfield{m}{\lambda^{-1}(s)}.
\end{construction}

\begin{proof}
 Let $\varphi = \bar{f} \circ \lambda$, 
i.e., the following diagram is commutative:
\[
\xymatrix{
  A \ar@{.>}[rr]^{f} \ar[d]_{\lambda} \ar[drr]^{\varphi}  
  &  & \bar{A} \ar@{.>}[d]^{\bar{\lambda}} \\
  S \ar[rr]^{\bar{f}}  &  &  \bar{S}.}
\]
Since $\bar{f}$ is \mfield{1}{S}, 
there is a unique $s \in \lambda(A)$ 
such that $\bar{f}(s) = \bar{s}$ for any 
$\bar{s} \in \varphi(A) = \bar{f}(\lambda(A))$.
Thus $\varphi^{-1}(\bar{s}) = \lambda^{-1}(s)$. 
Then the result follows from \cref{LocalLem}. 
\end{proof}

When $S = \bar{S}$ and $\bar{f}$ is the identity mapping,  
\cref{constr1} is reduced to \cref{LocalLem},
and thus they are equivalent to each other.
Moreover, \cref{constr1} generalizes 
\cite[Proposition~6]{2to1-MesQ19}
and \cite[Proposition~1]{GAO21mto1} 
which use the \ntoone definition, require that 
$\# A \equiv 0, 1 \pmod{n}$ and $\#S = \#\bar{S}$,
and only give the sufficient condition.

\subsection{The second construction}

\begin{construction}\label{constr2}
Let $A$, $\bar{A}$, $S$, $\bar{S}$ be finite 
sets and $f \colon A \rightarrow \bar{A}$, 
$\bar{f} \colon  S \rightarrow \bar{S}$, 
$\lambda \colon A \rightarrow S$, 
$\bar{\lambda} \colon \bar{A} \rightarrow \bar{S}$ 
be mappings such that 
$\bar{\lambda} \circ f = \bar{f} \circ \lambda$, 
i.e., the following diagram is commutative:
\[
\xymatrix{
  A \ar[rr]^{f}\ar[d]_{\lambda}  &   &  
  \bar{A}  \ar[d]^{\bar{\lambda}} \\
  S \ar[rr]^{\bar{f}}            &   &  \bar{S}.}
\]
Suppose~$\lambda$ is surjective,
$\# \lambda^{-1}(s) = m_1 \, 
\# \bar{\lambda}^{-1}(\bar{f}(s))$,  
and $f$ is \mfield{m_1}{\lambda^{-1}(s)} 
for any $s \in S$ and a fixed $m_1 \in \n$, where
\begin{align*}\label{eq:lambda-1s}
  \lambda^{-1}(s) & \coloneqq \{a \in A : \lambda(a) = s\}, \\    
  \bar{\lambda}^{-1}(\bar{f}(s)) & \coloneqq 
  \{b \in \bar{A} : \bar{\lambda}(b) = \bar{f}(s)\}.
\end{align*}
Then for $1 \le m \le m_1 \, \#S$, $f$ is $m$-to-$1$ on~$A$ \ifa 
$m_1 \mid m$, $\bar{f}$ is \mset{(m/m_1)}{S}, and
\begin{equation}\label{eq:SumPreim(s)=r}
    \sum_{s \in E_{\bar f}(S)} 
    \# \lambda^{-1}(s) = \# A \bmod{m},
\end{equation}
where $E_{\bar f}(S)$ is the exceptional set of $\bar{f}$ 
being \mset{(m/m_1)}{S}.
\end{construction}

\begin{proof}
Since $\lambda \colon A \rightarrow S$ is surjective, 
$A = \uplus_{s \in S} \lambda^{-1}(s)$ and 
\begin{align*}
\# A 
& = \sum_{s \in S} \# \lambda^{-1}(s)  
 = \sum_{s \in S} m_1 \, 
  \# \bar{\lambda}^{-1}(\bar{f}(s)) \\
& \ge \sum_{s \in S} m_1 = m_1 \# S.
\end{align*}
Thus,  the definitions that $f$ is $m$-to-$1$ on~$A$
and $\bar{f}$ is \mset{(m/m_1)}{S} are meaningful 
when $1 \le m \le m_1 \, \#S$.
For any $s \in S$, it follows from 
$\bar{\lambda} \circ f = \bar{f} \circ \lambda$ that
\begin{equation*}\label{eq:lamdfpreim(s)=f1s}
  (\bar{\lambda} \circ f) (\lambda^{-1}(s))
 = (\bar{f} \circ \lambda) (\lambda^{-1}(s))
 = \bar{f}(s),
\end{equation*}
and so $f(\lambda^{-1}(s)) \subseteq \bar{\lambda}^{-1}(\bar{f}(s))$.
Because 
\[
\# \lambda^{-1}(s) = m_1 \,\# \bar{\lambda}^{-1}(\bar{f}(s))
\]
and $f$ is \mfield{m_1}{\lambda^{-1}(s)}, we have
\[
\# f(\lambda^{-1}(s)) 
= \# \lambda^{-1}(s) / m_1 
= \# \bar{\lambda}^{-1}(\bar{f}(s)).
\]
Therefore,
\begin{equation}\label{eq:fpreim(s)=preim(fs)}
    f(\lambda^{-1}(s)) = \bar{\lambda}^{-1}(\bar{f}(s))
    \quad\text{for each $s \in S$.}
\end{equation}

Let $\varphi = \bar{\lambda} \circ f 
= \bar{f} \circ \lambda$, 
i.e., the following diagram is commutative:
\[
\xymatrix{
  A \ar[rr]^{f} \ar[d]_{\lambda} 
  \ar[drr]^{\varphi}  &  & 
  \bar{A} \ar[d]^{\bar{\lambda}} \\
  S \ar[rr]^{\bar{f}}  &  &  \bar{S}.}
\]
By \cref{LocalLem}, 
$f$ is $m$-to-$1$ on~$A$ \ifa $f$ is 
\mfield{m}{\varphi^{-1}(\bar{s})} 
for any $\# \varphi^{-1}(\bar{s}) \ge m$ and 
\begin{equation}\label{sumEf(subset)+sum=r}
    \sum_{\# \varphi^{-1}(\bar{s}) \ge m} 
    \# E_{f}(\varphi^{-1}(\bar{s})) 
  + \sum_{\# \varphi^{-1}(\bar{s}) < m} 
    \# \varphi^{-1}(\bar{s})
  = \#A \bmod{m},
\end{equation}
where $\bar{s}$ runs through $\varphi(A)$.
According to the proof of \cref{LocalLem}, 
if $f$ is $m$-to-$1$ on~$A$, then
\begin{equation}\label{Ef(subset)+set=Ef}
\Big( \underset{\# \varphi^{-1}(\bar{s}) \ge m}{\uplus}    
E_{f}(\varphi^{-1}(\bar{s})) \Big)
\uplus 
\Big( \underset{\# \varphi^{-1}(\bar{s}) < m}{\uplus}
    \varphi^{-1}(\bar{s}) \Big)
= E_f(A).
\end{equation}

For any $\bar{s} \in \varphi(A)$, assume  
there are exactly $m_{\bar{s}}$ distinct elements 
$s_1, s_2, \ldots, s_{m_{\bar{s}}} \in S$ such that 
\begin{equation}\label{eq:fs1=fsm}
    \bar{f}(s_1) 
  = \bar{f}(s_2) 
  = \cdots 
  = \bar{f}(s_{m_{\bar{s}}}) 
  = \bar{s},
\end{equation}
i.e., the set of preimages of $\bar{s}$ 
under~$\bar{f}$ is 
$\bar{f}^{-1}(\bar{s}) 
= \{ s_1, s_2, \ldots, s_{m_{\bar{s}}} \}$.
Then by \cref{eq:fpreim(s)=preim(fs)}, 
\begin{equation}\label{eq:flamdasi=barlamdabars}
  f(\lambda^{-1}(s_i)) 
= \bar{\lambda}^{-1}(\bar{f}(s_i)) 
= \bar{\lambda}^{-1}(\bar{s})
\end{equation}
for any $1 \le i \le m_{\bar{s}}$.
For any $s' \in S \setminus \bar{f}^{-1}(\bar{s})$,
we have $\bar{f}(s') \ne \bar{f}(s_1)$ and so
\begin{equation}\label{fsicaplamdas=0}
\begin{aligned}
  \varnothing
& = \bar{\lambda}^{-1}(\bar{f}(s')) \cap 
  \bar{\lambda}^{-1}(\bar{f}(s_1)) \\
& = f(\lambda^{-1}(s')) \cap f(\lambda^{-1}(s_1))  \\
& = f(\lambda^{-1}(s')) \cap \bar{\lambda}^{-1}(\bar{s}).
\end{aligned}
\end{equation}
It follows from $\varphi = \bar{f} \circ \lambda$ 
and \cref{eq:fs1=fsm} that
\begin{equation}\label{pre_bars=sum_pre_si}
  \varphi^{-1}(\bar{s}) 
  =  \lambda^{-1}(s_1) \uplus 
     \lambda^{-1}(s_2) \uplus\cdots \uplus
     \lambda^{-1}(s_{m_{\bar{s}}}).
\end{equation}
Then by \cref{eq:flamdasi=barlamdabars},
\[
  f(\varphi^{-1}(\bar{s})) 
  = f(\lambda^{-1}(s_1)) \cup\cdots\cup
    f(\lambda^{-1}(s_{m_{\bar{s}}}))
  = \bar{\lambda}^{-1}(\bar{s}).
\]
Since $A = \uplus_{s \in S} \lambda^{-1}(s)$,
$S = \bar{f}^{-1}(\bar{s}) \cup 
(S \setminus \bar{f}^{-1}(\bar{s}))$,
and \cref{fsicaplamdas=0} holds, 
it follows that the preimage set of 
$\bar{\lambda}^{-1}(\bar{s})$ under $f$ 
is $\varphi^{-1}(\bar{s})$. Because
\begin{equation}\label{eq:pre_si=m1pre_bars}
\#\lambda^{-1}(s_i) 
= m_1 \# \bar{\lambda}^{-1}(\bar{f}(s_i))
= m_1 \# \bar{\lambda}^{-1}(\bar{s})
\end{equation}
and $f$ is $m_1$-to-$1$ from $\lambda^{-1}(s_i)$ to 
\[
f(\lambda^{-1}(s_i)) 
= \bar{\lambda}^{-1}(\bar{f}(s_i)) 
= \bar{\lambda}^{-1}(\bar{s})
\]
for $1 \le i \le m_{\bar{s}}$, we get 
\begin{equation}\label{eq:numPre(bars)phi}
\begin{aligned}
& \# \varphi^{-1}(\bar{s}) 
= m_1 m_{\bar{s}} \# \bar{\lambda}^{-1}(\bar{s})
\text{~~and} \\ 
& \text{$f$ is $m_1 m_{\bar{s}}$-to-$1$ 
from $\varphi^{-1}(\bar{s})$ onto 
$\bar{\lambda}^{-1}(\bar{s})$.}
\end{aligned}
\end{equation}

We first prove the sufficiency. 
Suppose that $m_1 \mid m$, $\bar{f}$ is \mset{m_2}{S}, 
and \cref{eq:SumPreim(s)=r} holds, 
where $m_2 = m/m_1$. Define 
\begin{align*}
B_1 & = \{ \bar{f}(s) : s \in S \setminus E_{\bar{f}}(S)\}, \\
B_2 & = \{ \bar{f}(s) : s \in E_{\bar{f}}(S)\}.
\end{align*}
Then $\varphi(A) = B_1 \uplus B_2$.
When $\bar{s} \in B_1$, 
since $\bar{f}$ is \mfield{m_2}{S}, 
we have $\# \bar{f}^{-1}(\bar{s}) = m_2$.
By~\cref{eq:numPre(bars)phi}, 
\begin{equation}\label{num(phis)gem}
\# \varphi^{-1}(\bar{s}) 
= m_1 m_2 \# \bar{\lambda}^{-1}(\bar{s})
= m \# \bar{\lambda}^{-1}(\bar{s})
\ge m
\end{equation}
and $f$ is \mtoone from $\varphi^{-1}(\bar{s})$ 
onto $\bar{\lambda}^{-1}(\bar{s})$. Thus, 
\begin{equation}\label{sum=0ForsinB1}
   \sum_{\bar{s} \in B_1} 
   \# E_{f}(\varphi^{-1}(\bar{s})) 
   = 0.
\end{equation}
When $\bar{s} \in B_2$, we get 
$E_{\bar{f}}(S) = \uplus_{\bar{s} \in B_2} 
\bar{f}^{-1}(\bar{s})$. Then by
\cref{pre_bars=sum_pre_si} and 
\cref{eq:SumPreim(s)=r},
\begin{equation}\label{sum=rForsinB2}
\begin{aligned}
    \sum_{\bar{s} \in B_2} \# \varphi^{-1}(\bar{s})
    & = \sum_{\bar{s} \in B_2}
    \sum_{s_i \in \bar{f}^{-1}(\bar{s})} 
    \# \lambda^{-1}(s_i)  \\
    & = \sum_{s_i \in E_{\bar{f}}(S)} \# \lambda^{-1}(s_i) \\
    & = \# A \bmod{m} < m.
\end{aligned}
\end{equation}
Equations \cref{num(phis)gem}, 
\cref{sum=0ForsinB1}, and \cref{sum=rForsinB2} 
imply \cref{sumEf(subset)+sum=r}.
Then the sufficiency follows from \cref{LocalLem}.

We next prove the necessity. 
Assume $f$ is $m$-to-$1$ on~$A$. Let 
\begin{align*}
C_1 & = \{ \bar{s} \in \varphi(A) : 
\# \varphi^{-1}(\bar{s}) \ge m\}, \\
C_2 & = \{ \bar{s} \in \varphi(A) : 
\# \varphi^{-1}(\bar{s}) < m\}.
\end{align*}
Then $\varphi(A) = C_1 \uplus C_2$ and 
$S = \uplus_{\bar{s} \in \varphi(A)} \bar{f}^{-1}(\bar{s})$.
For any $\bar{s} \in C_1$, by \cref{LocalLem} 
and \cref{eq:numPre(bars)phi}, 
we obtain some equivalent statements:
(a) $f$ is \mfield{m}{\varphi^{-1}(\bar{s})};
(b) $m = m_1 m_{\bar{s}}$;
(c) $m_1 \mid m$ and $m_{\bar{s}} = m / m_1$;
(d) $m_1 \mid m$, $\bar{f}$ is 
\mset{(m/m_1)}{\bar{f}^{-1}(\bar{s})}, and 
$E_{\bar{f}}(\bar{f}^{-1}(\bar{s})) = \varnothing$.
We also note that $\# \varphi^{-1}(\bar{s}) 
= m \# \bar{\lambda}^{-1}(\bar{s})$. Thus,  
\begin{equation}\label{sum=0ForsinC1}
   \underset{\bar{s} \in C_1}{\uplus}  
   E_{f}(\varphi^{-1}(\bar{s})) 
   = \varnothing.
\end{equation}
Then \cref{Ef(subset)+set=Ef} minus 
\cref{sum=0ForsinC1} gives
\begin{equation}\label{EfC1=EfA}
  \underset{\bar{s} \in C_2}{\uplus}  
   \varphi^{-1}(\bar{s}) = E_f(A),  
\end{equation}
Substituting \cref{eq:numPre(bars)phi} 
into \cref{EfC1=EfA} yields  
\begin{equation*} 
  \sum_{\bar{s} \in C_2} 
    m_1 \# \bar{f}^{-1}(\bar{s}) 
    \# \bar{\lambda}^{-1}(\bar{s}) = r, 
\end{equation*}
where $r = \#A \bmod m < m$. Hence,
\begin{equation*} 
  \sum_{\bar{s} \in C_2} 
  \# \bar{f}^{-1}(\bar{s}) 
  \le r/m_1 < m/m_1.  
\end{equation*}
Combining this inequality and (d) yields that 
$m_1 \mid m$, $\bar{f}$ is 
\mset{(m/m_1)}{\uplus_{\bar{s} \in \varphi(A)} 
\bar{f}^{-1}(\bar{s}) = S}, and 
\[
E_{\bar{f}}(S) = \uplus_{\bar{s} 
\in C_2} \bar{f}^{-1}(\bar{s}). 
\]
Then by \cref{EfC1=EfA} and 
$\varphi = \bar{f} \circ \lambda$, we have
\begin{equation}\label{eq:EfA=sumEbarf(S)}
\begin{aligned}
E_f(A)
& = \underset{\bar{s} \in C_2}{\uplus}  
    \varphi^{-1}(\bar{s}) \\
& = \underset{\bar{s} \in C_2}{\uplus}
    \underset{s_i \in \bar{f}^{-1}(\bar{s})}{\uplus} 
    \lambda^{-1}(s_i)  \\
& = \underset{s_i \in E_{\bar{f}}(S)}{\uplus}
    \lambda^{-1}(s_i)
\end{aligned}
\end{equation}
and thus \cref{eq:SumPreim(s)=r} holds.
\end{proof} 

The identity \cref{eq:fpreim(s)=preim(fs)} 
plays an important role in the above proof.
Using this identity, the fact that 
$f$ is $m$-to-$1$ on~$A$ is divided into two parts: 
$f$ is \mfield{m_1}{\lambda^{-1}(s)} 
and $\bar{f}$ is \mfield{(m/m_1)}{S}.
When the first part holds, 
the problem whether $f$ is $m$-to-$1$ on~$A$  
is converted into that 
whether $\bar{f}$ is \mfield{(m/m_1)}{S}. 
The identity \cref{eq:SumPreim(s)=r} establishes a numerical relationship between the exceptional sets $E_{f}(A)$ and $E_{\bar{f}}(S)$. Moreover, this relationship is strengthened to \cref{eq:EfA=sumEbarf(S)} when $f$ is $m$-to-$1$ on~$A$.

The significance of \cref{constr2} 
resides in the fact that it not only 
unifies and generalizes the constructions in 
\cite{2to1-YuanZW21,nto1-NiuLQL23,Qin2469}
but also  provides more new explicit constructions. 

Applying \cref{constr2} to $m_1 = 1$ or $m \mid m_1 \, \#S$ 
yields the following results.

\begin{corollary}\label{mto1:m1=1}
Let $A$, $\bar{A}$, $S$, $\bar{S}$ be finite sets and
$f \colon A \rightarrow \bar{A}$, $\bar{f} \colon S \rightarrow \bar{S}$, 
$\lambda \colon A \rightarrow S$, $\bar{\lambda} \colon \bar{A} \rightarrow \bar{S}$ 
be mappings such that 
$\bar{\lambda} \circ f = \bar{f} \circ \lambda$. 
Suppose~$\lambda$ is surjective, $\# \lambda^{-1}(s) = \# \bar{\lambda}^{-1}(\bar{f}(s))$, 
and $f$ is \mfield{1}{\lambda^{-1}(s)} for any $s \in S$. 
Then for $1 \le m \le \#S$, 
$f$ is $m$-to-$1$ on~$A$ \ifa $\bar{f}$ is \mfield{m}{S} and
\[
\sum_{s \in E_{\bar f}(S)} \# \lambda^{-1}(s) = \#A \bmod{m}.
\]
\end{corollary}

\cref{mto1:m1=1} is a generalization of \cite[Theorem~4.3]{nto1-NiuLQL23} 
which uses the \ntoone definition, 
requires $\# A \equiv \# S \pmod{n}$, 
and does not give a necessary and 
sufficient condition in the case $n \nmid \# S$.

\begin{corollary}\label{mto1_m2|S}
With the notation and the hypotheses of \cref{constr2}, 
we assume $m \mid m_1 \, \#S$. Then $f$ is $m$-to-$1$ on~$A$ \ifa 
$m_1 \mid m$ and $\bar{f}$ is \mset{(m/m_1)}{S}.
\end{corollary}
\begin{proof}
We need only show \cref{eq:SumPreim(s)=r} holds 
when $m_1 \mid m$ and $\bar{f}$ is \mset{(m/m_1)}{S}.
In this case, $(m/m_1) \mid \# S$ and so 
$E_{\bar{f}}(S) = \varnothing$, which is equivalent to
$\sum_{s \in E_{\bar f}(S)} \# \lambda^{-1}(s) = 0$.
Note that $A = \uplus_{\bar{s} \in \varphi(A)}
\varphi^{-1}(\bar{s})$ and $\# \varphi^{-1}(\bar{s}) 
= m \,\# \bar{\lambda}^{-1}(\bar{s})$ for any 
$\bar{s} \in \varphi(A)$ by \cref{num(phis)gem}.
Thus $m \mid \# A$ and so \cref{eq:SumPreim(s)=r} holds.
\end{proof}

\cref{mto1_m2|S} generalizes 
{\cite[Proposition~4.2]{2to1-YuanZW21}} 
where $m=2$, $m_1 = 1$, $\#S$ is even, 
and only the sufficient condition is given.  
\cref{mto1_m2|S} reduces to the following form when $m = m_1$.

\begin{corollary}\label{mto1:m1=m} 
Let $A$, $\bar{A}$, $S$, $\bar{S}$ be finite sets and
$f \colon A \rightarrow \bar{A}$, 
$\bar{f} \colon S \rightarrow \bar{S}$, 
$\lambda \colon A \rightarrow S$, 
$\bar{\lambda} \colon \bar{A} \rightarrow \bar{S}$ 
be mappings such that 
$\bar{\lambda} \circ f = \bar{f} \circ \lambda$.
Suppose~$\lambda$ is surjective, 
$\# \lambda^{-1}(s) = m_1 \, \# \bar{\lambda}^{-1}(\bar{f}(s))$, 
and $f$ is \mfield{m_1}{\lambda^{-1}(s)} 
for any $s \in S$ and a fixed $m_1 \in \n$.
Then $f$ is \mset{m_1}{A} \ifa $\bar{f}$ is \mfield{1}{S}.
\end{corollary}

\cref{constr1} reduces to the sufficiency part of \cref{mto1:m1=m}
under the conditions that $\lambda$ is surjective,
$\# \lambda^{-1}(s) = m \, \# \bar{\lambda}^{-1}(\bar{f}(s))$, 
and $f$ is \mfield{m}{\lambda^{-1}(s)} 
for any $s \in S$ and a fixed $m \in \n$.
Moreover, \cref{mto1:m1=m} is a corrected version of \cite[Theorem~3.1]{Qin2469}.

\subsection{The third construction}

\begin{construction}\label{constr3}
Let $(G, \ast)$ be a finite group and 
$A$, $S$ be subsets of $G$. 
Let $f \colon A \rightarrow G$, 
$\bar{f} \colon S \rightarrow G$, 
$\lambda \colon A \rightarrow S$, 
$\bar{\lambda} \colon G \rightarrow G$
be mappings such that   
$\bar{\lambda} \circ f = \bar{f} \circ \lambda$, 
i.e., the following diagram is commutative:
\[
\xymatrix{
  A \ar[rr]^{f}\ar[d]_{\lambda}  &   &  
  G  \ar[d]^{\bar{\lambda}} \\
  S \ar[rr]^{\bar{f}}     &   &  G.}
\]
Assume $\bar{\lambda}$ is an endomorphism of~$G$ 
and $u$ is a mapping from~$A$ to~$G$ such that 
$\bar{\lambda}( u(a) )= c$ 
for any $a \in A$ and a fixed $c \in G$.
Let $f \ast u$ be the mapping defined by 
$f(a) \ast u(a)$ for $a \in A$. In particular,  
$(f \ast c) (a) = f(a) \ast c(a) = f(a) \ast c$. 
\begin{enumerate}[\upshape(1)]
\item \label{item:con3:1}
Suppose $\bar{f}$ is \mfield{1}{S} 
and $u = v \circ \lambda$, 
where $v$ is a mapping from~$S$ to~$G$. 
Then $f \ast u$ is $m$-to-$1$ on~$A$ \ifa 
$f$ is $m$-to-$1$ on~$A$, where $1 \le m \le \#A$.
\item \label{item:con3:2}
Suppose $\lambda$ is surjective,
$\# \lambda^{-1}(s) 
= m_1 \, \# \bar{\lambda}^{-1}(\bar{f}(s))$,  
and both~$f$ and $f \ast u$ are 
\mfield{m_1}{\lambda^{-1}(s)} 
for any $s \in S$ and a fixed $m_1 \in \n$. 
Then $f \ast u$ is $m$-to-$1$ on~$A$ \ifa 
$f$ is $m$-to-$1$ on~$A$, where $1 \le m \le m_1 \#S$.
\end{enumerate}
\end{construction}

\begin{proof}
Since $\bar{\lambda}$ is an endomorphism of~$G$, 
$\bar{\lambda}( u(a) )= c$, and 
$\bar{\lambda} \circ f = \bar{f} \circ \lambda$, we have
\begin{align*}
\bar{\lambda} \circ (f \ast u) 
 = (\bar{\lambda} \circ f) \ast 
    (\bar{\lambda} \circ u) 
 = (\bar{f} \circ \lambda) \ast c 
 = (\bar{f} \ast c) \circ \lambda,
\end{align*}
i.e., the following diagram is commutative:
\[
\xymatrix{
  A \ar[rr]^{f \ast u}\ar[d]_{\lambda}  &   &  
  G  \ar[d]^{\bar{\lambda}} \\
  S \ar[rr]^{\bar{f} \ast c}   &   &  G.}
\]

We first prove \cref{item:con3:1}. 
Clearly, $I \ast c$ permutes $G$,
where~$I$ is the identity mapping on~$G$. 
Also note that 
$\bar{f}$ maps $S$ to $G$ and 
$\bar{f} \ast c = (I \ast c) \circ \bar{f}$.
Hence, $\bar{f} \ast c$ is \mset{1}{S} 
\ifa $\bar{f}$ is \mset{1}{S}.  
By \cref{constr1}, 
$f \ast u$ is $m$-to-$1$ on~$A$ \ifa $f \ast u$ is \mfield{m}{\lambda^{-1}(s)} for any 
$\# \lambda^{-1}(s) \ge m$ and 
\[
  \sum_{\# \lambda^{-1}(s) \ge m} \# E_{f \ast u}(\lambda^{-1}(s)) 
  + \sum_{\# \lambda^{-1}(s) < m} \# \lambda^{-1}(s)
  = \#A \bmod{m}.
\]
By \cref{constr1}, $f$ is $m$-to-$1$ on~$A$ 
\ifa $f$ is \mfield{m}{\lambda^{-1}(s)} 
for any $\# \lambda^{-1}(s) \ge m$ and 
\[
  \sum_{\# \lambda^{-1}(s) \ge m} \# E_{f}(\lambda^{-1}(s)) 
  + \sum_{\# \lambda^{-1}(s) < m} \# \lambda^{-1}(s)
  = \#A \bmod{m}.
\]
For any $a \in \lambda^{-1}(s)$, 
i.e., $\lambda(a) = s$, we get
$u(a) = v (\lambda(a)) = v(s)$ and so
\begin{equation}\label{F=f+vs}
(f \ast u)(a) = f(a) \ast u(a) = f(a) \ast v(s).
\end{equation}
Hence, $f \ast u$ is \mfield{m}{\lambda^{-1}(s)} \ifa 
$f$ is \mfield{m}{\lambda^{-1}(s)}, and  
$E_{f \ast u}(\lambda^{-1}(s)) = E_{f}(\lambda^{-1}(s))$. 
Thus $f \ast u$ is $m$-to-$1$ on~$A$ \ifa $f$ is $m$-to-$1$ on~$A$.

We now prove \cref{item:con3:2}.
Since $\bar{\lambda}$ is an endomorphism of~$G$,  
\[
\# \bar{\lambda}^{-1}(\bar{f}(s) \ast c)
= \# \ker(\bar{\lambda})
= \# \bar{\lambda}^{-1}(\bar{f}(s))
\]
by \cref{EndKer},
and so $\# \lambda^{-1}(s) 
= m_1 \, \# \bar{\lambda}^{-1}(\bar{f}(s) \ast c)$.
Also note that $\lambda$ is surjective and 
$f \ast u$ is \mfield{m_1}{\lambda^{-1}(s)}.
Thus, by \cref{constr2}, $f \ast u$ is $m$-to-$1$ on~$A$ \ifa 
$m_1 \mid m$, $\bar{f} \ast c$ is \mset{m_2}{S}, and
\begin{equation}\label{eq:F+c=r}
    \sum_{s \in E_{\bar{f} \ast c}(S)} 
    \# \lambda^{-1}(s) = \# A \bmod{m},
\end{equation}
where $m_2 = m/m_1 \le \#S$.
By \cref{constr2} again, $f$ is $m$-to-$1$ on~$A$ 
\ifa $m_1 \mid m$, $\bar{f}$ is \mset{m_2}{S}, and
\begin{equation}\label{eq:f+c=r}
    \sum_{s \in E_{\bar f}(S)} 
    \# \lambda^{-1}(s) = \# A \bmod{m},
\end{equation}
where $m_2 = m/m_1 \le \#S$. Note that 
$\bar{f}$ maps $S$ to $G$, 
$I \ast c$ permutes $G$,
and $\bar{f} \ast c = (I \ast c) \circ \bar{f}$.
Hence, $\bar{f} \ast c$ is \mset{m_2}{S} 
\ifa $\bar{f}$ is \mset{m_2}{S}, and  
$E_{\bar{f} \ast c}(S) = E_{\bar{f}}(S)$, 
i.e., \cref{eq:F+c=r} is equivalent to \cref{eq:f+c=r}. 
Thus, $f \ast u$ is $m$-to-$1$ on~$A$ \ifa $f$ is $m$-to-$1$ on~$A$.
\end{proof}

This result reduces the problem whether 
$f \ast u$ is \mtoone on~$A$  
to that whether $f$ is \mtoone on~$A$. 
Thus, it provides a method for constructing 
new \mtoone mapping $f \ast u$ 
from known \mtoone mapping $f$
under certain conditions.

\begin{remark}
When $u = v \circ \lambda$, \cref{F=f+vs} implies 
that $f$ is \mfield{m_1}{\lambda^{-1}(s)} \ifa 
$f \ast u$ is \mfield{m_1}{\lambda^{-1}(s)}. 
Thus \cref{item:con3:2} also holds 
without the restriction that 
$f \ast u$ is \mfield{m_1}{\lambda^{-1}(s)} 
if $u = v \circ \lambda$. 
However, \cref{F=xmf} is in the case 
$u \ne v \circ \lambda$ of \cref{constr3}.
\end{remark}

\begin{remark}
When $(G, \ast) = (\fq, +)$, $c = 0$ and $m_1 = m = 1$,
\cref{item:con3:2} of \cref{constr3} is reduced to \cite[Theorem~3.2]{YD-AGW2}.
\end{remark}

\section{\texorpdfstring{\mtoone}{m-to-1} mappings of the form \texorpdfstring{$x^r h(x^s)$}{x\textasciicircum rh(x\textasciicircum s)}
}\label{sec_xrhxs}

In the rest of the paper, we consider only the \mtoone 
mappings of the form $x^r h(x^s)$ over finite fields. 
We first recall the well-known \onetoone property of such polynomials.

\begin{theorem}\label{xrhxs1to1}
Let $q-1 = \ell s$ for some $\ell, s \in \n$ and $h \in \fqx$.
Then $x^r h(x^s)$ permutes $\fq$ \ifa $(r, s) = 1$ 
and $x^r h(x)^s$ permutes $U_{\ell}$.
\end{theorem}

This result appeared in different forms in many 
references such as \cite{WanLidl91,ParkL01,AW07,Wang07,Zieve09}.
Many classes of permutation polynomials are constructed 
via an application of this result. 

For simplicity, we consider only the case that 
$x^r h(x^s)$ has only the root $0$ in $\fq$.
The following \mtoone relationship between 
$\fq$ and $\fqstar$ is a consequence of \cref{defn:mto1}.

\begin{lemma}\label{fq=fqstar}
Assume $f \in \fqx$ has only the root $0$ in $\fq$.
Then $f$ is \mfield{1}{\fq} \ifa $f$ is \mfield{1}{\fqstar}.
If $2 \le m \le q$, then $f$ is \mfq \ifa $m \nmid q$ and 
$f$ is \mfqstar. 
\end{lemma}
\begin{proof}
The first part is obvious. 
Assume $2 \le m \le q$ and $q = k m + r$, where $0 \le r \le m-1$. 
If $f$ is \mfq, then $0 \in E_f(\fq)$ and thus $r \ge 1$. 
Hence, $m \nmid q$ and $f$ is \mfqstar with 
$E_f(\fqstar) = E_f(\fq) \setminus \{0\}$.
If $m \nmid q$ and $f$ is \mfqstar, then $r \ne 0$ and so
$\# E_f(\fqstar) = (q-1) \bmod{m} \le m-2$.
Thus, $f$ is \mfq with 
$E_f(\fq) =  E_f(\fqstar) \cup \{0\}$.
\end{proof}

It is easy to check whether $m$ divides $q$.
Hence, for the polynomial $f$ having only the root $0$ in $\fq$,
we only consider the problem when~$f$ is \mfqstar in the sequel.
In particular, we obtain the following main theorem for $f(x) = x^r h(x^s)$.  

\begin{theorem}\label{MainThm}
Let $q-1 = \ell s$ and $m_1 = (r, s)$, 
where $\ell, r, s \in \n$.
Let $f(x) = x^r h(x^s)$ and $g(x) = x^{r_1} h(x)^{s_1}$, 
where $r_1 = r / m_1$, $s_1 = s / m_1$, 
and $h \in \fqx$ has no roots in 
$U_{\ell} \coloneqq \{\alpha \in \fqstar : \alpha^\ell = 1\}$.
If $1 \le m \le \ell m_1$, then $f$ is \mfqstar \ifa $m_1 \mid m$,
$g$ is \mfield{m_2}{U_{\ell}}, and $s (\ell \bmod{m_2}) < m$, 
where $m_2 = m / m_1$.
\end{theorem}
\begin{proof}
Evidently, 
$
  x^{s_1} \circ f 
  = x^{r s_1} h(x^s)^{s_1} 
  = x^{r_1 s} h(x^s)^{s_1} 
  = g \circ x^s 
$.
Since $\fqstar$ is a cyclic group and $s \mid q-1$, 
$x^s$ is $s$-to-$1$ from $\fqstar$ onto $U_{\ell}$. 
Because $h$ has no roots in $U_{\ell}$, 
$h(x^s) \ne 0$ for any $x \in \fqstar$, and so 
$f(\fqstar) \subseteq \fqstar$. 
Since $s_1 \mid q-1$, $x^{s_1}$ is $s_1$-to-$1$ 
from $\fqstar$ onto $U_{\ell m_1}$.
For any $\alpha \in U_{\ell}$, 
\[
  g(\alpha)^{\ell m_1} 
= \alpha^{r_1 \ell m_1} h(\alpha)^{s_1 \ell m_1}
= (\alpha^{\ell})^{r_1  m_1} h(\alpha)^{\ell s}
= 1
\]
and so $g(U_{\ell}) \subseteq U_{\ell m_1}$.
Thus the following diagram is commutative:
\[
\xymatrix{
  \fqstar \ar[rr]^{f}\ar[d]_{x^s}  &   &  
  \fqstar \ar[d]^{x^{s_1}} \\
  U_{\ell} \ar[rr]^{g}            &   &  U_{\ell m_1}.}
\]
Put $\lambda = x^s$ and $\bar{\lambda} = x^{s_1}$. 
It follows from $s = m_1 s_1$ that
$\# \lambda^{-1}(\alpha) = 
m_1 \# \bar{\lambda}^{-1}(g(\alpha))$
for any $\alpha \in U_{\ell}$. 
Write $\alpha = \xi^{is}$ for $\alpha \in U_{\ell}$, 
where $1 \le i \le \ell$ 
and $\xi$ is a primitive element of $\fq$.
Then $\lambda^{-1}(\alpha) 
= \xi^i \langle \xi^{\ell} \rangle$,
where $\langle \xi^{\ell} \rangle$
is a cyclic group of order~$s$.
Thus $f$ is \mfield{m_1}{\lambda^{-1}(\alpha)} 
by $(r, s) = m_1$.
According to \cref{constr2}, 
for $1 \le m \le m_1 \# U_{\ell}$,
$f$ is \mfqstar \ifa $m_1 \mid m$,
$g$ is \mfield{m_2}{U_{\ell}}, and  
\begin{equation}\label{sumEgUL=Fqmodm}
  \sum_{\alpha \in E_{g}(U_{\ell})} 
    \# \lambda^{-1}(\alpha) 
  = \# \fqstar \bmod m. 
\end{equation}
Let $\ell = \ell_2 m_2 + t$ with $0 \le t < m_2$. 
Then $\# E_{g}(U_{\ell}) = t$ and 
\[
q-1 = \ell s = \ell_2 s m_2 + s t 
= \ell_2 (s/m_1) m + s t.
\]
Hence, the right-hand side of \cref{sumEgUL=Fqmodm} 
is $st \bmod m$.
Since $\lambda$ is $s$-to-$1$ from $\fqstar$ onto $U_{\ell}$,
the left-hand side of \cref{sumEgUL=Fqmodm} is $st$.
Now \cref{sumEgUL=Fqmodm} becomes  
$st = st \bmod m$, i.e., $st < m$.
\end{proof}

From the above proof, we see that
\cref{MainThm} is a special case of \cref{constr2},
and the explicit condition $s (\ell \bmod{m_2}) < m$
is a simplified version of the restriction 
\cref{sumEgUL=Fqmodm} about exceptional sets.
\cref{MainThm} gives us a recipe 
to construct \mtoone mappings on $\fqstar$
from $m_2$-to-$1$ mappings on its subgroup $U_{\ell}$ 
under suitable conditions.

\begin{example}
Let $f(x) = x^2 h(x^4)$ and $g(x) = x h(x)^2$,
where $h(x) = x^2 + x + 3 \in \f_{17}[x]$.
Then~$U_4 = \{\pm 1, \pm 4\}$,
$h$ has no roots in $U_4$, 
$g$ is \mfield{3}{U_4}, and the 
exceptional set $E_g(U_4) = \{-4\}$.
Thus $f$ is \mfield{6}{\f_{17}^*}. 
Moreover, it is easy to verify that
$E_f(\f_{17}^*) = \{\pm 3, \pm 5\} 
= \lambda^{-1}(E_g(U_4))$, 
where $\lambda = x^4$.
\end{example}

\begin{remark}\label{hNoRoots}
In \cref{MainThm}, when $m < s \le \ell m_1$, 
the condition that $h$ has no roots in $U_{\ell}$
is necessary for~$f$ to be \mfqstar. 
Indeed, if~$h$ has a root in $U_{\ell}$, then by the fact 
that $x^s$ is $s$-to-$1$ from $\fqstar$ onto $U_{\ell}$,
we have $\# f^{-1}(0) \geq s > m$, which contradicts 
the definition of \mtoone.
\end{remark}

When $m = 1$, \cref{MainThm} 
is reduced to \cref{xrhxs1to1}. Moreover, 
applying \cref{MainThm} to $m_1 = 1$ 
or $m_2 = 1$ yields the following results.

\begin{corollary}\label{xrh(xs):m1=1}
Let $q-1 = \ell s$ and $(r, s) = 1$, where $\ell, r, s \in \n$.
Let $f(x) = x^r h(x^s)$ and $g(x) = x^{r} h(x)^{s}$, 
where $h \in \fqx$ has no roots in $U_{\ell}$.
Then $f$ is \mfqstar \ifa 
$g$ is \mfield{m}{U_{\ell}} and $s (\ell \bmod{m}) < m$, 
where $1 \le m \le \ell$.
\end{corollary}

\cref{xrh(xs):m1=1} generalizes 
 \cite[Proposition~4.9]{nto1-NiuLQL23} 
 where $m \mid \ell$.

\begin{corollary}\label{xrh(xs):m2=1}
Let $q-1 = \ell s$ and $m_1 = (r, s)$, 
where $\ell, r, s \in \n$.
Let $f(x) = x^r h(x^s)$ and $g(x) = x^{r_1} h(x)^{s_1}$, 
where $r_1 = r / m_1$, $s_1 = s / m_1$, 
and $h \in \fqx$ has no roots in $U_{\ell}$.
Then $f$ is \mfield{m_1}{\fqstar} 
\ifa $g$ is \mfield{1}{U_{\ell}}.
\end{corollary}

When $f(x) = x^r h(x^s)$, $\lambda (x) = x^s$, 
and $\bar{\lambda}(x) = x^{s_1}$,  
\cref{constr1} reduces to 
the sufficiency part of \cref{xrh(xs):m2=1}, 
and thus \cref{constr2} is more general than \cref{constr1}.
For this reason,  we will not consider \cref{constr1} in the sequel.

\cref{MainThm} converts the problem whether $f$ is \mfqstar  
to the second problem whether $g$ is \mfield{m_2}{U_{\ell}}.
In the following sections, we will study the second problem 
in depth for the special cases such as
\begin{enumerate}[\upshape(1)]
    \item $m = 2, 3$ or $\ell = 2, 3$;
    \item $g$ behaves like a monomial on $U_{\ell}$;
    \item $g$ behaves like a rational function on $U_{\ell}$;
    \item the second problem is further reduced  to another 
            problem by using \cref{constr2} again. 
\end{enumerate}
Similar results on the \onetoone mappings can be found 
in the survey paper~\cite{WangIndex19} and references therein.  

\section{Small \texorpdfstring{$m$ or $\ell$}{m or ell} }\label{sec_small_ml}


Applying \cref{MainThm} to $m=2$, $3$ yields the next results.

\begin{theorem}\label{xrh(xs):m=2}
Let $q-1 = \ell s$ and $m_1 = (r, s)$, 
where $r \ge 1$, $s \ge 2$, and $\ell \ge 2$.
Let $f(x) = x^r h(x^s)$ and $g(x) = x^{r_1} h(x)^{s_1}$, 
where $r_1 = r / m_1$, $s_1 = s / m_1$, 
and $h \in \fqx$ has no roots in $U_{\ell}$.
Then $f$ is \mfield{2}{\fqstar} \ifa one of the following holds:
\begin{enumerate}[\upshape(1)]
    \item $m_1 = 1$, $\ell$ is even, 
        and $g$ is \mfield{2}{U_{\ell}};
    \item \label{xrh(xs):m=2:2}
    $m_1=2$ and $g$ is \mfield{1}{U_{\ell}}.
\end{enumerate}
\end{theorem}

\begin{proof}
By \cref{MainThm}, $f$ is \mfield{2}{\fqstar} \ifa 
$m_1 \mid 2$, $s (\ell \bmod{m_2}) < 2$, and $g$ is 
\mfield{m_2}{U_{\ell}}, where $m_2 = 2 / m_1$. 
If $m_1 = 1$, then $m_2 = 2$. 
Since $s \ge 2$, $s (\ell \bmod{2}) < 2$ 
is equivalent to $2 \mid \ell$.
If $m_1 = 2$, then $m_2 = 1$ and 
$s (\ell \bmod{1}) = 0 < 2$. 
\end{proof}

\cref{xrh(xs):m=2:2} of \cref{xrh(xs):m=2} 
generalizes \cite[Proposition~16]{2to1-MesQ19} 
which only gives the sufficiency.
We next give an example of \cref{xrh(xs):m=2}.

\begin{corollary}
Let $f(x) = x^r (x^{\frac{2q-2}{3}} + x^{\frac{q-1}{3}} + a)$,
where $r \ge 1$, $q \ge 7$, $3 \mid q-1$, and $a \in \fq \setminus \{1, -2\}$. 
Then $f$ is \mfield{2}{\fqstar} \ifa 
$(r, \frac{q-1}{3}) = 2$, $r \equiv 2, 4 \pmod 6$, and 
$((a-1)^5(a+2))^{\frac{q-1}{6}} \notin \{\omega, \omega^2\}$, 
where $\omega$ is a primitive third root of unity over $\fq$.
\end{corollary}
\begin{proof}
Clearly, $\ell = 3$ and $U_3 = \{1, \omega, \omega^2\}$.
Let $h(x) = x^2 +x +a$. Then $h(1) = a+2$ and 
$h(\omega) = h(\omega^2) = a-1$, and so $h$ has no roots in $U_3$. 
By \cref{xrh(xs):m=2}, $f$ is \mfield{2}{\fqstar} \ifa $(r, \frac{q-1}{3}) = 2$  
and $g(x) \coloneqq x^{\frac{r}{2}} h(x)^{\frac{q-1}{6}}$ is \mfield{1}{U_3}, 
i.e., $g(1)$, $g(\omega)$, and $g(\omega^2)$ are distinct.    
The latter is equivalent to 
$((a-1)^5(a+2))^{\frac{q-1}{6}} \notin \{\omega^{\frac{r}{2}}, \omega^r\}$
and $\omega^{\frac{r}{2}} \ne 1$.
Then the result follows from $2 \mid r$ and $\ord(\omega) = 3$.
\end{proof}

\begin{theorem}\label{xrh(xs):m=3}
Let $q-1 = \ell s$ and $m_1 = (r, s)$, 
where $r \ge 1$, $s \ge 2$, and $\ell \ge 3$.
Let $f(x) = x^r h(x^s)$ and $g(x) = x^{r_1} h(x)^{s_1}$, 
where $r_1 = r / m_1$, $s_1 = s / m_1$, 
and $h \in \fqx$ has no roots in $U_{\ell}$.
Then $f$ is \mfield{3}{\fqstar} \ifa one of the following holds:
\begin{enumerate}[\upshape(1)]
    \item $m_1 = 1$, $\ell \equiv 0 \pmod{3}$, 
        and $g$ is \mfield{3}{U_{\ell}};
    \item $m_1 = 1$, $\ell \equiv 1 \pmod{3}$, $s = 2$, 
        and $g$ is \mfield{3}{U_{\ell}};
    \item $m_1=3$ and $g$ is \mfield{1}{U_{\ell}}.
\end{enumerate}
\end{theorem}

\begin{proof}
By \cref{MainThm}, 
$f$ is \mfield{3}{\fqstar} \ifa 
$m_1 \mid 3$, $s (\ell \bmod{m_2}) < 3$, 
and $g$ is \mfield{m_2}{U_{\ell}}, 
where $m_2 = 3 / m_1$. 
If $m_1 = 1$, then $m_2 = 3$. Since $s \ge 2$, 
$s (\ell \bmod{3}) < 3$ is equivalent to 
$\ell \equiv 0 \pmod{3}$ or
$\ell \equiv 1 \pmod{3}$ and $s=2$.
If $m_1 = 3$, then $m_2 = 1$ and 
$s (\ell \bmod{1}) = 0 < 3$.
\end{proof}


When $U_{\ell}$ has few elements, i.e.,~$\ell$ is small,
it is easy to determine the \mtoone property of~$g$ on $U_{\ell}$.
We demonstrate the cases $\ell = 2$, $3$ as an example.

\begin{theorem}\label{xrh(xs):L=2}
Let $q$ be odd, $s = (q-1)/2$, 
and $m_1 = (r, s)$, where $r, s \in \n$.
Let $f(x) = x^r h(x^s)$ and $g(x) = x^{r_1} h(x)^{s_1}$, 
where $r_1 = r / m_1$, $s_1 = s / m_1$, 
and $h \in \fqx$ with $h(1) h(-1) \ne 0$.
Then for $1 \le m \le 2 m_1$, $f$ is \mfqstar\ifa  
one of the following holds:
\begin{enumerate}[\upshape(1)]
    \item $m = m_1$ and $g(1) \ne g(-1)$;
    \item $m = 2 m_1$ and $g(1) = g(-1)$.
\end{enumerate}
\end{theorem}

Applying \cref{xrh(xs):L=2} to $h(x) = x + a$ 
yields the next~result.  

\begin{corollary} 
  Let $f(x) = x^r (x^{\frac{q-1}{2}} + a)$, where $r \in \n$, 
  $q$ is odd, and $a \in \fq \setminus \{\pm 1\}$.
  Then $f$ is \mfield{2}{\fqstar} \ifa one of the following holds:
\begin{enumerate}[\upshape(1)]
\item $(r, \frac{q-1}{2}) = 1$ and $(a^2-1)^{\frac{q-1}{2}} = (-1)^r$;
\item $(r, \frac{q-1}{2}) = 2$ and $((a+1)/(a-1))^{\frac{q-1}{4}} \ne (-1)^\frac{r}{2}$.
\end{enumerate} 
\end{corollary}
This result generalizes \cite[Theorem~4.14]{nto1-NiuLQL23} 
in which $q \equiv 3 \pmod{4}$ and $(r, \frac{q-1}{2}) = 1$.

\begin{theorem}\label{xrh(xs):L=3}
Let $s = (q-1)/3$ and $m_1 = (r, s)$, 
where $r$, $s \in \n$ and $3 \mid q-1$.
Let $f(x) = x^r h(x^s)$ and $g(x) = x^{r_1} h(x)^{s_1}$, 
where $r_1 = r / m_1$, $s_1 = s / m_1$, 
and $h \in \fqx$ has no roots in $U_3$.
Then for $1 \le m \le 3 m_1$, $f$ is \mfqstar
\ifa one of the following holds:
\begin{enumerate}[\upshape(1)]
\item $m = m_1$ and $g$ is \mfield{1}{U_3};
\item $m = 2 m_1$, $g$ is \mfield{2}{U_3},  and $s \mid r$;
\item $m = 3 m_1$ and $g$ is \mfield{3}{U_3}.
\end{enumerate}
\end{theorem}

Applying \cref{xrh(xs):L=3} to $h(x) = x - a$ 
yields the next~result.

\begin{corollary}\label{xrh(xs):L=3:h=x+a}
Let $f(x) = x^r (x^{\frac{q-1}{3}} - a)$
and $g(x) = x^{r_1} (x-a)^{s_1}$, 
where $r \in \n$, $3 \mid q-1$, 
$a \in \fq \setminus U_3$, 
$r_1 = r/(r, \frac{q-1}{3})$, 
and $s_1 = (q-1)/(3r, q-1)$.
Then $f$ is \mfield{3}{\fqstar} 
\ifa one of the following holds:
\begin{enumerate}[\upshape(1)]
    \item $(r, \frac{q-1}{3}) = 1$ and $g$ is \mfield{3}{U_3};
    \item $(r, \frac{q-1}{3}) = 3$ and $g$ is \mfield{1}{U_3}.
\end{enumerate}
\end{corollary}

\begin{example}
Let $f(x) = x^2(x^{21} + \xi^9)$ and $g(x) = x^2(x + \xi^9)^{21}$,
where $\xi$ is a primitive element of $\f_{64}$ such that 
$\xi^6 + \xi^4 + \xi^3 + \xi + 1 = 0$.
Then $g(1) = g(\omega) = g(\omega^2) = 1$, where $\omega = \xi^{21}$.
Hence $f$ is \mfield{3}{\f_{64}^{*}}. 
\end{example}

Similar results on the \onetoone mappings can be found 
in \cite{HFF,Hou15,WangIndex19} and references therein. 

\section{Monomials} 

The difficulty in applying \cref{MainThm} is to verify that $g$ is \mfield{m_2}{U_{\ell}}. However, the task is easy when $g$ behaves like a monomial on $U_{\ell}$. The results in this section are conjunctions of \cref{MainThm} and some results in \cite{AW07,Zieve09,Zieve130776}.  

\begin{theorem}
\label{xrh(xs):h(a)s1=bat}
Let $q-1 = \ell s$, $m_1 = (r, s)$, $r_1 = r / m_1$, and $s_1 = s / m_1$, where $\ell, r, s \in \n$. Let $h \in \fqx$ be such that $h(\alpha)^{s_1} = \beta \alpha^t$ for any $\alpha \in U_{\ell}$, a fixed $\beta \in U_{\ell}$, and a fixed integer $t$. Then $f(x) \coloneqq x^r h(x^s)$ is \mfqstar \ifa $m_1 \mid m$ and $(r_1 + t, \ell) = m / m_1$, where $1 \le m \le \ell m_1$.
\end{theorem}

\begin{proof}
For any $x \in U_{\ell}$, by $h(x)^{s_1} = \beta x^t$,  
we get $x^{r_1} h(x)^{s_1} = \beta x^{r_1 + t}$, 
which is \mfield{m_2}{U_{\ell}} \ifa 
$(r_1 + t, \ell) = m_2$. 
The result follows now from \cref{MainThm}.
\end{proof}

In \cref{xrh(xs):h(a)s1=bat}, 
$g(x) \coloneqq x^{r_1} h(x)^{s_1}$ 
behaves like the~monomial $\beta x^{r_1 + t}$ on $U_{\ell}$.
The following results give choices for the parameters 
satisfying the hypotheses of \cref{xrh(xs):h(a)s1=bat}.

\begin{corollary}
  Let $q-1 = \ell s$ and $m_1 = (r, s)$, 
  where $\ell$, $r$, $s \in \n$.
  Let $f(x) = x^r h(x^s)^{\ell m_1}$, 
  where $h \in \fqx$ has no roots in $U_{\ell}$.
  Then $f$ is \mfqstar \ifa $m_1 \mid m$ and 
  $(r, \ell m_1) = m$, where $1 \le m \le \ell m_1$.
\end{corollary}

\begin{proof}
For $\alpha \in U_{\ell}$,
$h(\alpha)^{\ell m_1 s_1} = h(\alpha)^{q-1} = 1$.
Now the result is in the special case $\beta = 1$ 
and $t = 0$ of \cref{xrh(xs):h(a)s1=bat}. 
\end{proof}

\subsection{\texorpdfstring{\mtoone mappings on $\fqtwostar$}{m-to-1 mappings on Fq2}}

Now we extend a class of permutations of 
$\fqtwostar$ in \cite[Theorem~5.1]{Zieve130776}
to \mtoone mappings on $\fqtwostar$.

\begin{theorem}\label{Mq-1=bxt} 
Suppose $M \in \fqtwox$ has no roots in 
$U_{q+1}$ and $\varepsilon x^t M(x)^q = M(x)$
for any $x \in U_{q+1}$, 
where $\varepsilon \in U_{q+1}$ 
and $\deg(M) \le t \le 2 \deg(M)$.
Let 
\[
f(x) = x^r M(x^{q-1})^{k m_1},
\]
where $r$, $k \in \n$ and $m_1 = (r, q - 1)$. 
Then $f$ is \mfqtwostar \ifa $m_1 \mid m$ 
and $(r_1 - k t, q + 1) = m/m_1$,
where $r_1 = r/m_1$ and $1 \le m \le m_1(q+1) $.
\end{theorem}
\begin{proof}
Since $0 \ne \varepsilon x^t M(x)^q = M(x)$
for $x \in U_{q+1}$, we get
$M(x)^{q-1} = \varepsilon^{-1} x^{-t}$,
and so $M(x)^{k(q-1)} = \varepsilon^{-k} x^{-k t}$.
Then the result follows from \cref{xrh(xs):h(a)s1=bat}.
\end{proof}

When $t = \deg(M)$ and $k = m_1 = m = 1$,
\cref{Mq-1=bxt} is equivalent to
\cite[Theorem~5.1]{Zieve130776}.

\begin{remark}
The polynomial $M$ satisfying 
$\varepsilon x^t M(x)^q = M(x)$
for any $x \in U_{q+1}$ can be determined explicitly. 
Indeed, let $M(x) = \sum_{i=0}^{d} a_i x^i \in \fqtwox$,
where $d = \deg(M)$. Then a direct computation gives that
$\varepsilon x^t M^q = M$ \ifa
\begin{equation}\label{eq:xtMq=M}
M(x) = \sum_{i=t-d}^{\lfloor t/2 \rfloor} 
(a_i x^i + \varepsilon a_i^q x^{t-i}),
\end{equation}
where $\lfloor t/2 \rfloor$ 
denotes the largest integer $\le t/2$.
\end{remark}

For simplicity, take $t = d$, $a_0 = -a \in U_{q+1}$, 
$\varepsilon = -a$, and other $a_i = 0$.
Then $M(x) = x^d - a$, and it has no roots in $U_{q+1}$ 
\ifa $a \notin (U_{q+1})^d$. 
Hence we obtain the following result.

\begin{corollary}\label{fq2:h=xt-a}
Let $a \in \fqtwo$ satisfy $a^{q+1} = 1$ and $a^t \ne 1$, 
where $t = (q+1)/(d, q+1)$ with $d \in \n$.
Let 
\[
f(x) = x^r(x^{d(q-1)} - a)^{k m_1},
\]
where $r$, $k \in \n$ and $m_1 = (r, q-1)$.  
Then $f$ is \mfqtwostar \ifa 
$m_1 \mid m$ and $(r_1-kd, q+1) = m/m_1$,
where $r_1=r/m_1$ and $1 \le m \le m_1(q+1)$.
\end{corollary}

\begin{example}\label{fq2:h=x4+1}
Let $q$ be odd such that $3 \mid q+1$ and $8 \nmid q+1$.
Then $x^{4q-3} + x$ is \mfield{3}{\fqtwostar}.
\end{example}

\begin{example}\label{fq2:h=x3-a}
Let $q$ be odd and $3 \mid q+1$. Let $a \in \fqtwo$ 
satisfy $a^{q+1} = 1$ and $a^{(q+1)/3} \ne 1$.
Then $x^{3q-2} - ax$ is \mfield{2}{\fqtwostar}.
\end{example}

\begin{example}
Let $q = 2^{10k + 5}$ with $k \ge 0$. Let $a \in \fqtwo$ 
satisfy $a^{q+1} = 1$ and $a^{(q+1)/11} \ne 1$.
Then $x^{22q-21} + a^2 x$ is \mfield{3}{\fqtwostar}.
\end{example}

\subsection{\texorpdfstring{\mtoone mappings on $\fqnstar$}{m-to-1 mappings on Fqn}}

Next we extend two classes of permutations of $\fqnstar$ 
in~\cite{Zieve09} to \mtoone mappings on $\fqnstar$.

\begin{theorem}\label{xrhxs_Fqn_q-1} 
 Let $q^n-1 = \ell s$, $m_1 = (r, s)$, and $\ell m_1 \mid (q-1, n)$,
 where $n$, $\ell$, $s$, $r \in \n$. 
 Let $f(x) = x^rh(x^s)$, where $h \in \fqx$ has no roots in $U_{\ell}$. 
 Then $f$ is \mfield{m_1}{\fqnstar}. 
\end{theorem}
\begin{proof}
Let $r_1 = r/m_1$ and $s_1 = s/m_1$. 
By $q \equiv 1 \pmod{\ell m_1}$, 
\[
 \frac{q^{\ell m_1} - 1}{q - 1} 
 = \sum^{\ell m_1 - 1}_{i = 0} q^i \equiv 0 \pmod{\ell m_1},
\]
and so $q - 1$ divides $(q^{\ell m_1} - 1)/(\ell m_1)$, 
which divides 
\[ (q^n - 1)/(\ell m_1) = s/m_1 = s_1.
\]
Thus $q-1 \mid s_1$. 
For $\alpha \in U_{\ell}$, we have $\alpha \in \fq^*$ by $\ell \mid q-1$, 
and so $h(\alpha) \in \fqstar$. 
Then $h(\alpha)^{s_1} = 1$ by $q-1 \mid s_1$. 
Since $\ell \mid q - 1$ and $q-1 \mid s_1$, we get $\ell \mid s_1$. 
Then $(r_1, \ell) = 1$ by $(r_1, s_1) = 1$. 
Now the result is in the special case $t = 0$ and $m = m_1$
of \cref{xrh(xs):h(a)s1=bat}. 
\end{proof}

In the following results we use the notation
\begin{equation}\label{hd}
h_d(x) = x^{d-1} + x^{d-2} + \cdots + x + 1.
\end{equation}

\begin{theorem}\label{xrhxs_Fqn_q+1}
  Let $q^n - 1 = \ell s$, $m_1 = (r, s)$, 
  and $\ell m_1 \mid q + 1$,  
  where $n$ is even, $\ell, s, r \in \n$. Assume 
  \[
  h(x) \coloneqq h_d(x^e)^t H(h_k(x^e)^{\ell_0})
  \]
  has no roots in $U_{\ell}$, where $H \in \fqx$,
  $d, e, t, k \in \n$, 
  and $\ell_0 = \ell/(\ell,k-1)$.
  Then for $1 \le m \le \ell m_1$, 
  $f(x) \coloneqq x^r h(x^{s})$ 
  is \mfield{m}{\fqnstar} \ifa $m_1 \mid m$ and 
  \begin{equation}\label{eq:r1+qk=m2}
      \Big(\ell m_1, r + \frac{(1 - d) e s t }{q-1} \Big) = m.
  \end{equation}
\end{theorem}

\begin{proof}
Since $n$ is even and $\ell m_1 \mid q + 1$, 
we have the divisibility relations
\[
q-1 = \frac{q^2-1}{q+1} \mid \frac{q^n-1}{q+1} 
\mid \frac{q^n-1}{\ell m_1} = s_1,
\]
where $s_1 = s/m_1$.
For $\alpha \in U_{\ell} \setminus \{1\}$,  we get 
$\alpha^{q} = \alpha^{-1}$ by $\ell \mid q+1$, and so 
\[
h_k(\alpha)^{q} 
= \Big( \frac{\alpha^k-1}{\alpha-1} \Big)^{q} 
= \frac{\alpha^{-k}-1}{\alpha^{-1}-1} 
= \frac{h_k(\alpha)}{\alpha^{k-1}}.
\]
Then $h_k(\alpha)^{\ell_0q} = h_k(\alpha)^{\ell_0}$, 
i.e., $h_k(\alpha)^{\ell_0} \in \fq$. 
Clearly, $h_k(1) = k \in \fq$. 
Since $H \in \fqx$ and $H(h_k(x^e)^{\ell_0})$ 
has no roots in $U_{\ell}$, we have 
$H(h_k(\alpha^{e})^{\ell_0}) \in \fqstar$
for any $\alpha \in U_{\ell}$.
Thus $H(h_k(\alpha^{e})^{\ell_0})^{s_1} = 1$.

 For $\alpha \in U_{\ell}$, if $\alpha^e \ne 1$,
 then $\alpha^{q} = \alpha^{-1}$ by $\ell \mid q+1$, and so
 \[
 h_d(\alpha^e)^{q} = \Big( \frac{\alpha^{ed}-1}{\alpha^e-1} \Big)^{q} = \frac{\alpha^{-ed}-1}{\alpha^{-e}-1} = \frac{h_d(\alpha^e)}{\alpha^{e(d-1)}}.
 \]
 By hypothesis, $h_d(x^e)$ has no roots in $U_{\ell}$,  
 and so $h_d(\alpha^e)^{q-1} = \alpha^{e(1-d)}$. Thus, 
\[
 h_d(\alpha^e)^{s_1} 
 = h_d(\alpha^e)^{(q-1)s_1/(q-1)} 
 = \alpha^{e(1-d)s_1/(q-1)}.
\] 
If $\alpha^e = 1$, then 
$h_d(\alpha^e)^{s_1} = h_d(1)^{s_1} = d^{s_1} = 1$ 
by $d \in \fqstar$ and $q-1 \mid s_1$.
Thus $h_d(\alpha^e)^{s_1} = \alpha^{e(1-d)s_1/(q-1)}$ 
for any $\alpha \in U_{\ell}$.
Then the result follows from \cref{xrh(xs):h(a)s1=bat}. 
\end{proof}

The following lemma characterizes the condition 
that $h_d(x^e)$ has no roots in $U_{\ell}$.

\begin{lemma}\label{lem:hdxeUL}
  Let $U_{\ell}$ be the cyclic group of 
  all $\ell$-th roots of unity over $\fqn$, 
  where $\ell, n \in \n$ and $\ell \mid q^n-1$.
  Then $h_d(x^e)$ has no roots in $U_{\ell}$ 
  \ifa $(d, q \ell/(e, \ell)) = 1$,
  where $d, e \in \n$.
\end{lemma}

\begin{proof}
  Evidently, $h_d(1) \ne 0$ \ifa $(d, q) = 1$. 
  For $\alpha \in U_{\ell} \setminus \{1\}$, 
  $h_d(\alpha) = (\alpha^d - 1)/(\alpha-1)$. 
  Then $h_d(\alpha) \ne 0$ \ifa $\alpha^d \ne 1$,
  which is equivalent to $(d, \ell) = 1$.
  Hence, $h_d(x)$ has no roots in $U_{\ell}$ 
  \ifa $(d, q \ell) = 1$.
  Note that $x^e$ is $(e, \ell)$-to-$1$ 
  from $U_{\ell}$ onto $U_{\ell/(e, \ell)}$.
  Thus, $h_d(x^e)$ has no roots in $U_{\ell}$ \ifa 
  $h_d(x)$ has no roots in $U_{\ell/(e, \ell)}$, 
  which is equivalent to $(d, q \ell/(e, \ell)) = 1$.
\end{proof}

Applying \cref{xrhxs_Fqn_q-1} to $h(x) = h_d(x^{e})^t$
and \cref{xrhxs_Fqn_q+1} to $H(x) = 1$
yields the following results.

\begin{corollary}\label{xrhxs_Fqn_q-1_cor}
   Let $q^n-1 = \ell s$, $m_1 = (r, s)$, 
   and $\ell m_1 \mid (q-1, n)$,  
   where $n, \ell, s, r \in \n$. 
   Let $f(x) = x^r h_d(x^{es})^t$, where $d, e, t \in \n$ 
   with $(d, q\ell/(e, \ell)) = 1$.
   Then $f$ is \mfield{m_1}{\fqnstar}.
\end{corollary}

\begin{corollary}\label{xrhxs_Fqn_q+1_cor}
  Let $q^n-1=\ell s$, $m_1=(r, s)$, and $\ell m_1 \mid q + 1$, 
  where $n$ is even, $\ell, s, r \in \n$.
  Let $f(x) = x^r h_d(x^{es})^t$, where $d, e, t \in \n$ 
   with $(d, q\ell/(e, \ell)) = 1$.
   Then $f$ is \mfield{m}{\fqnstar} \ifa $m_1 \mid m$ and 
   \cref{eq:r1+qk=m2} holds, where $1 \le m \le \ell m_1$.
\end{corollary}

The results in this subsection extend 
Theorems~1.2 and~1.3, Corollaries~2.3 and~2.4 in~\cite{Zieve09}
where $m_1 = 1$ and $(e, \ell) = 1$.

\section{Rational functions}

In this section, we demonstrate the case that~$g$ 
behaves like a rational function on $U_{\ell}$.
Part~1 presents a class of \mtoone mappings on  
$\fqtwostar$ by using a known \onetoone rational function.
Parts~2 and~3 give two classes of rational functions 
that are \thrtoone and \mfield{5}{U_{q+1}} respectively, 
by employing the decompositions of related algebraic curves.

Applying \cref{MainThm} to $\ell = q + 1$ 
and $s = q-1$ yields the next result.

\begin{theorem}\label{xrh(xs):fq2}
Let $f(x) = x^r h(x^{q-1})$ and $g(x) = x^{r_1} h(x)^{s_1}$, 
where $h \in \fqtwox$ has no roots in $U_{q+1}$,
$r \ge 1$, $r_1 = r / m_1$, $s_1 = (q-1) / m_1$, and $m_1 = (r, q-1)$.
Then $f$ is \mfield{m}{\fqtwostar} \ifa
$m_1 \mid m$, $g$ is \mfield{m_2}{U_{q+1}}, 
and $(q-1) (q+1 \bmod{m_2}) < m$, 
where $1 \le m \le m_1 (q+1)$ and $m_2 = m / m_1$.
\end{theorem}

\subsection{Known \texorpdfstring{\onetoone}{1-to-1} rational functions}

\begin{lemma}[\!\!{\cite[Lemma~2.2]{Gupta16}}]
\label{h(Uq+1)ne0}
For $n \in \n$, $x^4 + x + 1$ and $x^4 + x^3 + 1$ 
have no roots in $U_{2^n +1}$.
\end{lemma}

\begin{lemma}[\!\!{\cite[Lemma~3.2]{ZhaHF17}}]
\label{Uq+1:x4+x3+1}
Let $q = 2^n$ with $n \ge 1$. Then   
\[
  G(x) \coloneqq \frac{x^5 + x^2 + x}{x^4 + x^3 + 1} 
\]
permutes $U_{q+1}$ \ifa $n$ is even. 
\end{lemma}

\begin{theorem}\label{f1f2mto1}
Let $q = 2^n$ with $n$ even. 
Then $x^{4q + 1} + x^{3q + 2} + x^{5}$
is \mfield{(5, q-1)}{\fqtwostar}.
\end{theorem}
\begin{proof}
  Put $h(x) = x^4 + x^3 + 1$ and  
  $g(x) = x^{\frac{5}{m_1}} h(x)^{\frac{q-1}{m_1}}$, 
  where $m_1 = (5, q-1)$. Then
  \begin{align*}
    x^{m_1} \circ g(x) 
    & = x^5 h(x)^{q-1} \\
    & = \frac{x^5(x^4 + x^3 + 1)^q}{x^4 + x^3 + 1} \\
    & = \frac{x^5(x^{-4} + x^{-3} + 1)}{x^4 + x^3 + 1} \\
    & = G(x) 
  \end{align*}
  for $x \in U_{q+1}$.
  By \cref{Uq+1:x4+x3+1}, $G$ is \mfield{1}{U_{q+1}},
  and so $g$ is \mfield{1}{U_{q+1}}.
  Then the result follows form 
  \cref{h(Uq+1)ne0,xrh(xs):fq2}. 
\end{proof}

\cref{f1f2mto1} extends \cite[Theorems~3.1 and~3.2]{ZhaHF17}
where $n \equiv 2 \pmod{4}$.

\subsection{New \texorpdfstring{\thrtoone}{3-to-1} rational function}\label{sec:3to1}

We begin with a different proof of a result in \cite{Lachaud90,Dillon2004,KLi18,KimM2020}. 


\begin{lemma}[\cite{Lachaud90,Dillon2004,KLi18,KimM2020}]
\label{a+1/a:2to1:even}
Let 
\[
T_i = \{c \in \ftwonstar \mid \trtnt(1/c) = i\}
\]
with $i \in \{0, 1\}$. Then the mapping $a \mapsto a + a^{-1}$ 
is \twotoone from $\ftwon \setminus \{0, 1\}$ onto $T_0$ 
and it is also \twotoone from $U_{2^n+1} \setminus \{1\}$ onto $T_1$, 
where $U_{2^n+1} = \{a \in \f_{2^{2n}} \mid a^{2^n+1} = 1\}$.
\end{lemma}
\begin{proof}
For $a \in U_{2^n+1} \setminus \{1\}$, 
we have $a + 1/a \in \ftwonstar$ and 
\begin{align*}
    \trtnt(1/(a + 1/a))
    & = \trtnt(1/(a+1) + 1/(a^2+1)) \\
    & = 1/(a+1) + 1/(a^{2^n} + 1) \\
    & = 1,
\end{align*}
i.e., $a + 1/a \in T_1$.
For any $a$, $a+b \in U_{2^n+1} \setminus \{1\}$, 
if $a + 1/a = (a + b) + 1/(a+b)$, 
then $b = 0$ or $b = (a^2+1)/a \ne 0$. 
Thus, $a \mapsto a + 1/a$ is \twotoone 
from $U_{2^n+1} \setminus \{1\}$ onto $T_1$ 
with cardinality $2^{n-1}$.
Similarly, $a \mapsto a + 1/a$ is \twotoone  
from $\ftwon \setminus \{0, 1\}$ onto $T_0$ 
with cardinality $2^{n-1}-1$.
\end{proof}

\begin{corollary}\label{tr(1/c)}
For any $c \in \ftwonstar$, 
$\trtnt(1/c) = 0$ \ifa $c = a + a^{-1}$ 
for some $a \in \ftwon \setminus \{0, 1\}$,
and $\trtnt(1/c) = 1$ \ifa $c = a + a^{-1}$ 
for some $a \in U_{2^n+1} \setminus \{1\}$.
\end{corollary}

We next give a new class of \thrtoone rational functions on $U_{2^n+1}$.

\begin{theorem}\label{g3to1}
Let $c \in \ftwonstar$  with $n \ge 1$ and
\[
g(x) = \frac{c x^3 + x^2 + 1}{x^3 + x + c}.
\]
If $\trtnt(1 + c^{-1}) = 0$, 
then $g$ is \mfield{1}{U_{2^n+1}}.
If $\trtnt(1 + c^{-1}) = 1$, 
then $g$ is \mfield{3}{U_{2^n+1}}. 
\end{theorem}

\begin{proof}
Put $q = 2^n$. Proposition~3.1 (i) in 
{\cite{BartQ1889}} implies that 
$x^3 + x + c$ has no roots in $U_{q+1}$.
Then for any $x$, $y \in U_{q+1}$, 
$g(x) = g(y)$ is equivalent to
\begin{equation}\label{eq:g(x)=g(y)deg3}
    (c x^3 + x^2 + 1) (y^3 + y + c) 
  = (c y^3 + y^2 + 1) (x^3 + x + c).
\end{equation}
Proposition~3.2 (ii) in {\cite{BartQ1889}}
states that \cref{eq:g(x)=g(y)deg3} factors as 
\begin{equation}\label{eq:g(x)=g(y)Fq}
  (x + y) H_1(x, y) H_2(x, y) = 0, 
\end{equation}
where $H_1(x, y) =  x y + \alpha x + \beta y + 1$, 
$H_2(x, y) = x y + \beta x + \alpha y + 1$,
and $\alpha, \beta \in \fqtwo$ are the roots 
of $Q(x) \coloneqq x^2 + c x + c^2 + 1$. 
Thus $\alpha + \beta = c$ 
and $\alpha \beta = c^2 + 1$.

(i) When $\trtnt(1 + c^{-1}) = 0$, 
we get $\tr_{q/2}((c^2+1)/c^2) = 0$
and thus $\alpha, \beta \in \fq$. 
For any $x$, $y \in U_{q+1}$,  
\begin{align*}
x y H_1(x, y)^q 
& = x y (x y + \alpha x + \beta y + 1)^q \\
& = x y (x^{-1} y^{-1} + \alpha x^{-1} 
        + \beta y^{-1} + 1) \\
& = x y + \beta x + \alpha y + 1 \\
& = H_2(x, y) 
\end{align*}
and hence the roots of $H_1$ and $H_2$ are the same. 
For $x$, $y \in U_{q+1}$, if $H_1(x, y) \ne 0$, 
then $H_2(x, y) \ne 0$ and thus $x = y$ 
by \cref{eq:g(x)=g(y)Fq}, which implies that 
$g$ is \mfield{1}{U_{2^n+1}}.
Therefore, we need only show that 
if $H_1(x, y) = 0$, then $x = y$.

If $\beta \in U_{q+1}$, then 
$\beta \in \fq \cap U_{q+1} = \{1\}$, 
i.e., $\beta = 1$. Since $\alpha + \beta = c$ 
and $\alpha \beta = c^2 + 1$, 
we get $\alpha = c + 1 = c^2 + 1$. 
Hence $c = 1$ and so $\alpha = 0$. 
Then $H_1(x, y) =  x y + y + 1$. 
If $H_1(x, y) = 0$, then $x \ne 1$ 
and $y = (x+1)^{-1}$. By $y^q = y^{-1}$,
we get $x^2 + x = 1$ and thus $y = (x+1)^{-1} = x$.

If $\beta \notin U_{q+1}$, then $x + \beta \ne 0$
for any $x \in U_{q+1}$. If $H_1(x, y) = 0$, then
$y = (\alpha x + 1)/(x + \beta)$ and so
\[
y^q = \Big(\frac{\alpha x + 1}{x + \beta}\Big)^q
= \frac{\alpha x^{-1} + 1}{x^{-1} + \beta}
= \frac{\alpha + x}{1 + \beta x}.
\]
By $y \in U_{q+1}$ and $\alpha + \beta = c \ne 0$, 
we get the following equivalent statements:
\begin{align*}
y^q = y^{-1} 
~\Longleftrightarrow &~~ 
\frac{\alpha + x}{1 + \beta x} 
    = \frac{x + \beta}{\alpha x + 1}\\
\Longleftrightarrow &~~ 
(\alpha + \beta) x^2 + (\alpha + \beta)^2 x 
    + (\alpha + \beta) = 0\\
\Longleftrightarrow &~~ 
x^2 + (\alpha + \beta) x + 1 = 0 \\
\Longleftrightarrow &~~ 
x = (\alpha x + 1)/(x + \beta) = y.
\end{align*}

(ii) When $\trtnt(1 + c^{-1}) = 1$, we have 
\[
\tr_{q/2}((c^2+1)/c^2) = 1.
\]
Thus $Q$ is irreducible over $\fq$ and so 
$\alpha$, $\beta \in \fqtwo \setminus \fq$ 
with $\beta = \alpha^q$.
Since $\alpha^{q+1} = c^2 + 1$ and $c \ne 0$, 
we get $\alpha$, $\alpha^q \notin U_{q+1}$. Let
\begin{equation}\label{eq:ThreeRoots}
\begin{aligned}
  y_0 & = x, \\
  y_1 & = (\alpha x + 1)/(x + \alpha^q), \\
  y_2 & = (\alpha^q  x + 1)/(x + \alpha).
\end{aligned}
\end{equation}
Then for any $x \in U_{q+1}$,  
\[
 y_1^q 
  = \Big(\frac{\alpha x + 1}{x + \alpha^q}\Big)^q
  = \frac{x + \alpha^q}{\alpha x + 1}
  = \frac{1}{y_1}
\]
and $y_2^q = y_2^{-1}$ similarly.
Thus $(x, y_i) \in U_{q+1} \times U_{q+1}$, 
$0 \le i \le 2$, are solutions of \cref{eq:g(x)=g(y)Fq}.
Hence, $g$ is \mfield{3}{U_{q+1}} \ifa
$y_0$, $y_1$, $y_2$ are distinct for any $x \in U_{q+1}$ 
except for $(q+1) \bmod{3}$ elements.
By \cref{eq:ThreeRoots} and $\alpha + \alpha^q = c$, 
it is easy to verify that $y_i = y_j$ for any 
$i \ne j \in \{0, 1, 2\}$ \ifa $x^2 + c x + 1 = 0$.

If $n$ is odd, then $3 \mid q+1$ 
and $\tr_{q/2}(1) = 1$.
Since $\tr_{q/2}(1 + c^{-1}) = 1$,
we get $\tr_{q/2}(1/c) = 0$. 
By \cref{tr(1/c)}, $x^2 + c x + 1$ 
has two distinct roots $x_0$, $x_0^{-1}$ 
in $\fq \setminus \{0, 1\}$.
Hence for any $x \in U_{q+1}$, $x^2 + c x + 1 \ne 0$,  
and thus $y_0$, $y_1$, $y_2$ are distinct. 
Therefore, $g$ is \mfield{3}{U_{q+1}}.

If $n$ is even, then $q + 1 \equiv 2 \pmod{3}$
and $\tr_{q/2}(1) = 0$.
Since $\tr_{q/2}(1 + c^{-1}) = 1$,
we get $\tr_{q/2}(1/c) = 1$. 
By \cref{tr(1/c)}, $x^2 + c x + 1$ 
has two distinct roots $x_0$, $x_0^{-1}$ 
in $U_{q+1} \setminus \{1\}$. Hence for any 
$x \in U_{q+1} \setminus \{x_0, x_0^{-1}\}$,
$x^2 + c x + 1 \ne 0$, 
and thus $y_0$, $y_1$, $y_2$ are distinct.
Therefore, $g$ is \mfield{3}{U_{q+1}}.
\end{proof}

\cref{g3to1} can also be obtained from the perspective of composition of rational functions; see \cref{rem:g=hx3h}. Moreover, this result generalizes {\cite[Lemma~4.1]{ZhaHF17}} where $c=1$ and {\cite[Proposition~3.2 (ii)]{BartQ1889}} where $n$ is even and $\trtnt(1/c) = 0$. 
It also implies the following result.
 
\begin{theorem}\label{g1_3to1}
Let $g_1(x) = x (x^3 + x + c)^{\frac{2^n-1}{3}}$,
where $c \in \ftwonstar$ and $n$ is even.  
If $\tr_{2^n/2}(1/c) = 0$,
then $g_1$ is \mfield{1}{U_{2^n+1}}.
If $\tr_{2^n/2}(1/c) = 1$,
then $g_1$ is \mfield{3}{U_{2^n+1}}.
\end{theorem}

\begin{proof}
Let $q = 2^n$. For any $x \in U_{q+1}$, 
$x^q = x^{-1}$ and thus
\begin{equation}\label{eq:rf3_h=x3+x+c}
\begin{aligned}
x^3 \circ g_1 
& = \frac{x^3(x^3 + x + c)^q}{x^3 + x + c} \\
& = \frac{x^3(x^{-3} + x^{-1} + c)}{x^3 + x + c} \\
& = \frac{c x^3 + x^2 + 1}{x^3 + x + c}.
\end{aligned}    
\end{equation}
If $\tr_{q/2}(1/c) = 0$, then $x^3 \circ g_1$ 
is \mfield{1}{U_{q+1}} by \cref{g3to1},
and thus $g_1$ is \mfield{1}{U_{q+1}}.

If $\tr_{q/2}(1/c) = 1$, 
then $x^2 + c x + c^2 + 1$ has two roots 
$\alpha$, $\alpha^q \in \fqtwo \setminus \fq$,
where $\alpha + \alpha^q = c$ and
$\alpha^{q+1} = c^2 + 1$.
Thus, $\alpha$, $\alpha^q \notin U_{q+1}$,
$\alpha^q = \alpha + c$,  
$\alpha^2 = c \alpha + c^2 + 1$, and
$\alpha^3 = \alpha \alpha^2 = \alpha + c^3 + c$. 
Let  
\begin{align*}
  y_0 & = x, \\
  y_1 & = (\alpha x + 1)/(x + \alpha^q), \\ 
  y_2 & = (\alpha^q  x + 1)/(x + \alpha).
\end{align*}
In the proof of \cref{g3to1}, we have already shown that 
$y_0$, $y_1$, $y_2 \in U_{q+1}$ and they are distinct for any
$x \in U_{q+1} \setminus \{x_0, x_0^{-1}\}$, where
$x_0$, $x_0^{-1} \in U_{q+1}$ are the roots of $x^2 + c x + 1$. 
To prove that $g_1$ is \mfield{3}{U_{q+1}}, we need only show 
\[
g_1 (y_0) = g_1 (y_1) = g_1 (y_2)
\]
for any
$x \in U_{q+1} \setminus \{x_0, x_0^{-1}\}$. 
Indeed, for any $x \in U_{q+1}$, 
\begin{align*}
\frac{\alpha x + 1}{(x + \alpha^q)^q}  
= \frac{\alpha x + 1}{x^q + (\alpha + c)^q}  
= \frac{\alpha x + 1}{x^{-1} + \alpha}   
= x \quad\text{and}  
\end{align*}
\begin{align*}
    & \phantom{{}={}} 
    (\alpha x + 1)^3 + (\alpha x + 1)(x + \alpha^q)^2 + c(x + \alpha^q)^3 \\ 
    & = (\alpha x + 1)^3 + (\alpha x + 1)(x + \alpha + c)^2  
      + c(x + \alpha + c)^3 \\
    & =  (\alpha^3 + \alpha + c) x^3 
          + (\alpha^3 + c \alpha^2 + (c^2 + 1) \alpha + c^3) x \\
    & \phantom{{}={}}
        + c(\alpha + c)^3 + (\alpha + c)^2  + 1 \\
    & = c^3 x^3 + c^3 x + c^4.
\end{align*}
Therefore, \vspace{-9pt}
\begin{align*}
    g_1 (y_1)
    & = \frac{\alpha x + 1}{x + \alpha^q} 
        \Big(\Big(\frac{\alpha x + 1}{x + \alpha^q}\Big)^3 
                + \frac{\alpha x + 1}{x + \alpha^q} + c
            \Big)^{\frac{q-1}{3}} \\
    & = x (c^3 x^3 + c^3 x + c^4)^{\frac{q-1}{3}} \\ 
    & = x (x^3 + x + c)^{\frac{q-1}{3}} \\ 
    & = g_1 (y_0)  
\end{align*} \vspace{-3ex}
and $g_1 (y_2) = g_1 (y_0)$ by a similar argument.
\end{proof}

\begin{corollary}\label{f3to1_h=x3+x+c}
Let $f(x) = x^{3q} + x^{q+2} + c x^3$,
where $q = 2^n$ with $n \ge 2$ and $c \in \fqstar$.
Then $f$ is \mfield{1}{\fqtwostar} 
\ifa $n$ is odd and $\tr_{q/2}(1/c) = 1$,
and $f$ is \mfield{3}{\fqtwostar} 
\ifa $\tr_{q/2}(1/c) = 0$.
\end{corollary}

\begin{proof}
Fix $h(x) = x^3 + x + c$. Then $f(x) = x^3 h(x^{q-1})$.
Let $m_1 = (3, q-1)$ and $g(x) = x^{3/m_1} h(x)^{(q-1)/m_1}$.
By \cref{xrh(xs):fq2}, $f$ is \mfield{1}{\fqtwostar}
\ifa $m_1 = 1$ and $x^{3} h(x)^{q-1}$ is \mfield{1}{U_{q+1}}, 
i.e., $n$ is odd and $\tr_{q/2}(1/c) = 1$ 
by \cref{eq:rf3_h=x3+x+c} and \cref{g3to1}.

By \cref{xrh(xs):m=3}, $f$ is \mfield{3}{\fqtwostar}
\ifa (1) $m_1 = 1$, $3 \mid q+1$, 
and $x^{3} h(x)^{q-1}$ is \mfield{3}{U_{q+1}},
or (2) $m_1 = 3$ and $x h(x)^{(q-1)/3}$ 
is \mfield{1}{U_{q+1}}.
If $n$ is odd, then $m_1 = 1$ and $3 \mid q+1$.
By \cref{g3to1}, $x^{3} h(x)^{q-1}$ is 
\mfield{3}{U_{q+1}} \ifa $\tr_{q/2}(1/c) = 0$. 
Thus, $f$ is \mfield{3}{\fqtwostar} 
\ifa $\tr_{q/2}(1/c) = 0$.
If $n$ is even, then $m_1 = 3$.
By \cref{g1_3to1}, $x h(x)^{(q-1)/3}$ is 
\mfield{1}{U_{q+1}} \ifa $\tr_{q/2}(1/c) = 0$.
Thus, $f$ is \mfield{3}{\fqtwostar} 
\ifa $\tr_{q/2}(1/c) = 0$.
\end{proof}

The \onetoone part of \cref{f3to1_h=x3+x+c} is known;
see for example \cite{KLiQLC20,Ozbudak22,Ozbudak23cc}. 
The sufficiency of the \thrtoone part can also be verified 
from the perspective of composition of functions. 
Indeed, if $\trtnt(1/c) = 0$, then $c = a + a^{-1}$ for some 
$a \in \ftwon \setminus \{0, 1\}$ by \cref{tr(1/c)}. 
A direct computation yields that
$a \sigma \circ f \circ \sigma = (a + 1)^{6} x^3$,
where $\sigma(x) = x^q + a x$.
Since~$\sigma$ permutes $\fqtwostar$ and~$x^3$ 
is \mfield{3}{\fqtwostar}, it follows 
that~$f$ is \mfield{3}{\fqtwostar}.

A mapping $f$ from $\fq$ to itself is called 
differentially $d$-uniform (abbreviated $d$-uniform) if
\[
d = \max_{a, b \in \fq, a \neq 0}
\# \{ x \in \fq : f(x + a) - f(x) = b \}.
\]
In particular, $f$ is called almost perfect nonlinear 
(APN) if it is $2$-uniform.
Recently, K{\"{o}}lsch et~al.\ \cite{KolschKK23}
established the following equivalence relation between 
$d$-uniform mappings and $(d+1)$-to-$1$ mappings.


\begin{lemma}[\!\!{\cite[Corollary~7]{KolschKK23}}]
\label{APNand3to1}
Let $q = 2^n$ with $n$ even and 
\[
f(x) = \sum\limits_{0 \leq i < j \leq n - 1} 
a_{i j} x^{2^i + 2^j} \in \fqx.
\]
Assume that $f(x) = g(x^3)$ for some $g \in \fqx$. 
Then~$f$ is APN on $\fq$ \ifa $f$ is \mfield{3}{\fqstar}. 
\end{lemma}

Combining \cref{f3to1_h=x3+x+c,APNand3to1} yields the next result.

\begin{corollary}\label{x3qAPN}
Let $f(x) = x^{3q} + x^{q+2} + c x^3$,
where $q = 2^n$ with $n$ even and $c \in \fqstar$. 
Then~$f$ is APN on $\fqtwo$ \ifa $\tr_{q/2}(1/c) = 0$. 
\end{corollary}

This result gives an infinite class of APN trinomials,
but it is linear equivalent to the well known 
Gold function~$x^3$ by the remark after \cref{f3to1_h=x3+x+c}.

K{\"{o}}lsch et~al.\ \cite{KolschKK23} also showed that all APN functions on $\ftwon$ with the minimal image size are very close to being \thrtoone. Budaghyan et~al.\ \cite{BudaghyanIK2335} provided a list of all known quadratic \thrtoone APN functions on $\ftwon$ for $n \leq 12$ up to EA-equivalence.
Beierle et~al.\ \cite{Beierle26MillionsAPN} 
computationally constructed 3,775,599 CCZ-inequivalent 
quadratic APN functions over $\f_{2^8}$ 
and estimated the total number to be about 6 million. 
Among them, many APN functions meet the condition of
\cref{APNand3to1}, and thus they are \thrtoone. 
Therefore, it would be interesting to explore the method of 
constructing \mtoone mappings presented in this paper 
for finding new classes of quadratic APN functions of the form $g(x^3)$.

\subsection{New \texorpdfstring{$5$-to-$1$}{5-to-1} rational function}\label{sec:5to1}

\begin{theorem}\label{g5to1}
Let $q = 2^n$ with $n \ge 1$ and 
\[
  g(x) = \frac{x^4 + x + 1}{x^5 + x^4 + x}.
\]
If $n \equiv 2 \pmod{4}$, then $g$ is \mfield{5}{U_{q+1}}.
If $n \not\equiv 2 \pmod{4}$, then $g$ is \mfield{1}{U_{q+1}}.
\end{theorem}

\begin{proof}
By \cref{h(Uq+1)ne0}, 
$x(x^4 + x^3 + 1)$ has no roots in $U_{q+1}$.
Hence for any $x$, $y \in U_{q+1}$, 
$g(x) = g(y)$ is equivalent to
\begin{equation}\label{eq:g(x)=g(y)deg5}
    (x^4 + x + 1) (y^5 + y^4 + y) 
  = (x^5 + x^4 + x) (y^4 + y + 1).
\end{equation}
{\cite[Page~8]{BartG181}}
states that \cref{eq:g(x)=g(y)deg5} factors as 
\begin{equation}\label{eq:g(x)=g(y)F16}
  (x + y) \prod_{i=1}^{4} 
  (x y + \omega^{2^{i-1}} x 
    + \omega^{2^{i+1}} y + 1) = 0, 
\end{equation}
where $\omega$ is a primitive element of $\f_{16}$ 
such that $\omega^4 + \omega + 1 = 0$.
This factorization can be verified manually 
or by a computer program.
Since $\ord(\omega^{2^{i+1}}) = 15$ and 
$q+1 \not\equiv 0 \pmod{15}$,
we get $\omega^{2^{i+1}} \notin U_{q+1}$, and so 
$x + \omega^{2^{i+1}} \ne 0$ for any $x \in U_{q+1}$, 
where $1 \le i \le 4$. Let
\begin{equation}\label{eq:FiveRoots}
  y_0 = x  \quad\text{and}\quad
  y_i = (\omega^{2^{i-1}} x + 1)/(x + \omega^{2^{i+1}}), 
\end{equation}
where $1 \le i \le 4$.
Then $(x, y_i) \in U_{q+1} \times K$, $0 \le i \le 4$, 
are solutions of \cref{eq:g(x)=g(y)F16},  
where $K$ is an extension field of $\fqtwo$.
Thus, $g$ is \mfield{1}{U_{q+1}} \ifa
\cref{eq:g(x)=g(y)F16} has no roots 
$(x, y) \in U_{q+1}^2$ with $x \ne y$.
When $5 \mid q+1$, $g$ is \mfield{5}{U_{q+1}} \ifa
$y_0$, $y_1$, \ldots, $y_4$ in $U_{q+1}$ and
they are distinct for any $x \in U_{q+1}$.

For $1 \le i \le 4$, a direct computation yields that  
$y_i^q = 1/y_i$ \ifa $\alpha x^2 + \beta x + \gamma = 0$, where 
\begin{align*}
  \alpha & = \omega^{2^{i-1}} + \omega^{2^{i+1} q}, \\
  \beta & = \omega^{2^{i-1}} \omega^{2^{i-1} q} 
        + \omega^{2^{i+1}}\omega^{2^{i+1} q}, \\
  \gamma & = \omega^{2^{i-1} q} + \omega^{2^{i+1}}.
\end{align*}
Because
\begin{equation}\label{eq:omega2i=1Fq2}
  \omega^{2^{i-1}} + \omega^{2^{i+1}} 
= (\omega + \omega^4)^{2^{i-1}} = 1,
\end{equation}
we have 
\begin{align*}
\beta 
& = \omega^{2^{i-1}} (\omega^{2^{i+1}} + 1)^q 
    + (\omega^{2^{i-1}} + 1) \omega^{2^{i+1} q} 
= \alpha, \\
\gamma & = (\omega^{2^{i+1}} + 1)^q 
+ (\omega^{2^{i-1}} + 1) = \alpha.
\end{align*}
Thus $y_i \in U_{q+1}$ \ifa $\alpha (x^2 + x + 1) = 0$.
Since $\omega^{16} = \omega$, 
$\alpha = 0$ \ifa $n \equiv 2 \pmod{4}$.

If $n \equiv 2 \pmod{4}$, then $\alpha = 0$
and so $y_i \in U_{q+1}$ for  $1 \le i \le 4$.
Hence \cref{eq:g(x)=g(y)F16} has five solutions 
$y_0$, $y_1$, \ldots, $y_4$ in $U_{q+1}$ for any $x \in U_{q+1}$. 
(i) Assume $y_i = y_0$ for some $i \in \{1, 2, 3, 4\}$.
By \cref{eq:FiveRoots,eq:omega2i=1Fq2}, 
$y_i = y_0$ is equivalent to $x^2 + x + 1 = 0$. 
Thus $x^3 = 1$ and $x \ne 1$, 
a contradiction to that $U_{q+1}$ has 
no elements of order~$3$ by $(3, q+1) = 1$. 
(ii) Assume $y_i = y_j$ for some $i \ne j \in \{1, 2, 3, 4\}$. 
By \cref{eq:FiveRoots}, $y_i = y_j$ is equivalent to
\begin{multline}\label{eq:yi=yjUq+1}
  (\omega^{2^{i-1}} + \omega^{2^{j-1}}) x^2 
  + (\omega^{2^{i-1}} \omega^{2^{j+1}} 
    + \omega^{2^{i+1}} \omega^{2^{j-1}}) x \\
  + \omega^{2^{i+1}} + \omega^{2^{j+1}} =0,
\end{multline}
Since $\ord(\omega) = 15$, 
$\omega^{2^{i-1}} \ne \omega^{2^{j-1}}$ 
for $i \ne j \in \{1, 2, 3, 4\}$. 
By \cref{eq:omega2i=1Fq2}, $\omega^{2^{i+1}} = \omega^{2^{i-1}} + 1$.
Hence, \cref{eq:yi=yjUq+1} is equivalent to $x^2 + x + 1 =0$, 
a contradiction to that $U_{q+1}$ has no elements of order~$3$. 
Combining (i) and (ii), we see that $y_0$, $y_1$, \ldots, $y_4$ are distinct.
Note that $5 \mid q+1$. Therefore, $g$ is \mfield{5}{U_{q+1}}.

If $n \not \equiv 2 \pmod{4}$, then $\alpha \ne 0$.
Hence $y_i \in U_{q+1}$ for $i \in \{1, 2, 3, 4\}$
\ifa $x^2 + x + 1 = 0$.
When $n \equiv 0 \pmod{4}$, we have $(3, q+1) = 1$,
and so $U_{q+1}$ has no elements of order~$3$.
Thus, $y_i \notin U_{q+1}$ for any $i \in \{1, 2, 3, 4\}$,
i.e., \cref{eq:g(x)=g(y)F16} has no roots 
$(x, y) \in U_{q+1}^2$ with $x \ne y$.
When $n \equiv 1, 3 \pmod{4}$, we get $3 \mid q+1$, 
and so $U_{q+1}$ has two elements of order~$3$.
Then $y_i \in U_{q+1}$, i.e., $x^2 + x + 1 = 0$, implies that
\[
  \omega^{2^{i-1}} x + 1 
= \omega^{2^{i-1}} x + x + x^2
= x(\omega^{2^{i+1}} + x),
\]
i.e., $y_i = x$ for any $i \in \{1, 2, 3, 4\}$ by \cref{eq:FiveRoots}.
Hence, \cref{eq:g(x)=g(y)F16} also has no roots 
$(x, y) \in U_{q+1}^2$ with $x \ne y$.
Therefore, $g$ is \mfield{1}{U_{q+1}} 
if $n \not \equiv 2 \pmod{4}$.
\end{proof}

\cref{g5to1} extends some results in
\cite{Gupta16,NLiH17_triF2n,KLi18_tri}
which only consider the \onetoone property 
of~$g$ by using different methods. 
Moreover, \cref{g5to1} can also be obtained 
from the perspective of composition of 
rational functions; see \cref{rem:g=hx3h}.

\begin{corollary}\label{f3to1_h=x4+x3+1}
Let $f(x) = x^{4q-1} + x^{3q} + x^{3}$,
where $q = 2^n$ with $n \ge 2$. 
Then $f$ is \mfield{1}{\fqtwostar} \ifa $n$ is odd, and  
$f$ is \mfield{3}{\fqtwostar} \ifa $n \equiv 0 \pmod{4}$.
\end{corollary}

\begin{proof}
Fix $h(x) = x^4 + x^3 + 1$. 
Then $h$ has no roots in $U_{q+1}$ 
by \cref{h(Uq+1)ne0} and $f(x) = x^3 h(x^{q-1})$.
For any $x \in U_{q+1}$, $x^q = x^{-1}$ and so 
\begin{align*}
  x^3 h(x)^{q-1} 
  & = \frac{x^3(x^4 + x^3 + 1)^q}{x^4 + x^3 + 1} \\
  & = \frac{x^3(x^{-4} + x^{-3} + 1)}{x^4 + x^3 + 1} \\
  & = \frac{x^4 + x + 1}{x^5 + x^4 + x}.
\end{align*}
Let $m_1 = (3, q-1)$ and $g(x) = x^{3/m_1} h(x)^{(q-1)/m_1}$.
By \cref{xrh(xs):fq2}, $f$ is \mfield{1}{\fqtwostar}
\ifa $m_1 = 1$ and $x^{3} h(x)^{q-1}$ is \mfield{1}{U_{q+1}}, 
i.e., $n$ is odd by \cref{g5to1}.

\cref{g5to1} implies $x^{3} h(x)^{q-1}$ is not 
\mfield{3}{U_{q+1}}. Thus, by \cref{xrh(xs):m=3}, 
$f$ is \mfield{3}{\fqtwostar} \ifa $m_1 = 3$ and 
$g_1(x) \coloneqq x h(x)^{(q-1)/3}$ is \mfield{1}{U_{q+1}}. 
The condition $m_1 = 3$ is equivalent to~$n$ is even.
If $n \equiv 0 \pmod{4}$, then 
$x^3 \circ g_1$ is \mfield{1}{U_{q+1}} by \cref{g5to1},
and so $g_1$ is \mfield{1}{U_{q+1}}.
If $n \equiv 2 \pmod{4}$, then 
$x^3 \circ g_1$ is \mfield{5}{U_{q+1}} by \cref{g5to1}.
Since $g_1$ induces a map from $U_{q+1}$ to $U_{3(q+1)}$
and $x^3$ is a \thrtoone map from $U_{3(q+1)}$ to $U_{q+1}$,
we have $g_1$ is not \mset{1}{U_{q+1}}.
Hence, $f$ is \mfield{3}{\fqtwostar} \ifa $n \equiv 0 \pmod{4}$. 
\end{proof}

The \onetoone part of \cref{f3to1_h=x4+x3+1} 
is the same as \cite[Theorem 3.5]{Gupta16}.
We remark that the polynomial~$f$ in 
\cref{f3to1_h=x4+x3+1} can be written as 
$f(x) = x^{4q-1} + \tr_{q^2/q}(x^3)$,
and permutation polynomials of more general form
$x^s + \gamma \trqnq(x^t)$ over $\fqn$ 
are intensively studied; see for example 
\cite{GoharP08Planar,Charpin09,Gohar11,Charpin14,
GerikeK19,KLiQCL18,JiangYLQ25xtrq2q,JiangLQ26xtrq3q}.

\section{Further reduction}\label{sec_further_reduction}

By using \cref{constr2} repeatedly, the following result converts the second problem whether~$g$ is \mfield{m_2}{U_{\ell}} to the third problem whether~$\bar{g}$ is $m_2/m_3$-to-$1$ on another set~$S$.

\begin{theorem}\label{xrh(xs)2}
Let $q-1 = \ell s$ and $m_1 = (r, s)$, 
where $\ell$, $r$, $s \in \n$.
Let $f(x) = x^r h(x^s)$ and 
$g(x) = x^{r_1} h(x)^{s_1}$, 
where $r_1 = r / m_1$, $s_1 = s / m_1$, 
and $h \in \fqx$ has no roots in $U_{\ell}$.
Let $S$, $\bar{S}$ be finite sets and 
$\lambda \colon U_{\ell} \rightarrow S$, 
$\bar{\lambda} \colon U_{\ell m_1} \rightarrow \bar{S}$, 
$\bar{g} \colon S \rightarrow \bar{S}$
be mappings such that~$\lambda$ is surjective and 
$\bar{\lambda} \circ g = \bar{g} \circ \lambda$.
That is, the following diagrams are commutative:
\[
\xymatrix{
  \fqstar \ar[rr]^{f}\ar[d]_{x^s}  &   &  
  \fqstar \ar[d]^{x^{s_1}} \\
  U_{\ell} \ar[rr]^{g}\ar[d]_{\lambda}  &   &  
  U_{\ell m_1}  \ar[d]^{\bar{\lambda}} \\
  S \ar[rr]^{\bar{g}}    &   &  \bar{S}.
  }
\]
Suppose $\# \lambda^{-1}(\alpha) = 
m_3 \, \# \bar{\lambda}^{-1}(\bar{g}(\alpha))$ 
and $g$ is \mfield{m_3}{\lambda^{-1}(\alpha)}
for any $\alpha \in S$ and a fixed $m_3 \in \n$.
Then $f$ is \mfqstar \ifa $m_1 m_3 \mid m$,
$s (\ell \bmod{m_2}) < m$, $\bar{g}$ is 
\mfield{m/(m_1 m_3)}{S}, and 
\begin{equation}\label{eq:SumPreim(s)=Lmodm2}
    \sum_{\alpha \in E_{\bar{g}}(S)} 
    \# \lambda^{-1}(\alpha) = \ell \bmod{m_2},
\end{equation}
where $1 \le m \le m_1 m_3 \, \# S$, $m_2 = m / m_1$,
and $E_{\bar{g}}(S)$ is the exceptional set of $\bar{g}$ 
being \mfield{m/(m_1 m_3)}{S}.
\end{theorem}

\begin{proof}
By \cref{MainThm}, $f$ is \mfqstar \ifa $m_1 \mid m$,
$g$ is \mfield{m_2}{U_{\ell}}, and $s (\ell \bmod{m_2}) < m$,
where $1 \le m \le \ell m_1$.
Thus, $f$ is not \mfqstar if $1 \le m < m_1$, 
i.e., the result holds when $1 \le m < m_1$.
Applying \cref{constr2} to 
the lower commutative diagram yields that
$g$ is \mfield{m_2}{U_{\ell}} \ifa $m_3 \mid m_2$,
$\bar{g}$ is \mfield{(m_2/m_3)}{S}, and 
\cref{eq:SumPreim(s)=Lmodm2} holds,
where $1 \le m_2 \le m_3 \, \# S$.
Note that $m_1 m_3 \mid m$ is equivalent to 
$m_1 \mid m$ and $m_3 \mid m_2$. 
Since~$\lambda$ is surjective,  
\begin{align*}
\ell & = \# U_{\ell}
= \sum_{\alpha \in S} \# \lambda^{-1}(\alpha) \\
& = \sum_{\alpha \in S} m_3 \, 
    \# \bar{\lambda}^{-1}(\bar{g}(\alpha))  \\
& \ge m_3 \, \# S.
\end{align*}
The conditions $1 \le m \le \ell m_1$ and 
$1 \le m_2 \le m_3 \, \# S$ imply that 
$m_1 \le m \le m_1 m_3 \, \# S$.
Thus the result holds when $m_1 \le m \le m_1 m_3 \, \# S$.
This completes the proof. 
\end{proof}

To simplify the construction of commutative diagrams, 
assume $f(x) = x^r H(x^{q-1})^{m_1} \in \fqtwox$, 
where $m_1 = (r, q-1)$. 
Then $g(x) = x^{r/m_1} H(x)^{q-1}$ 
and it maps $U_{q+1}$ to $U_{q+1}$.
To further reduce the problem, 
we mainly consider two cases:
\begin{enumerate}[\upshape(1)]
\item $\lambda$ and $\bar{\lambda}$ are \onetoone 
    from $U_{q+1}$ to $U_{q+1}$ and $\bar{g} = x^n$;
\item $\lambda$ and $\bar{\lambda}$ are \onetoone 
    from $U_{q+1}$ to $\fq \cup \{\infty\}$ 
    and $\bar{g} = x^n$.
\end{enumerate}

\subsection{\texorpdfstring{$\lambda$ is \onetoone from $U_{q+1}$ to itself}{lambda is 1-to-1 from Uq+1 to itself}}

\begin{theorem}\label{xrhxs(m1)_Uq+1}
Let $L_1, L_2, M_1, M_2 \in \fqtwox$ satisfy 
that $M_i$ has no roots in $U_{q+1}$,
$L_i = \varepsilon_i x^{t_i} M_i^q$ 
for any $x \in U_{q+1}$, 
and $L_i / M_i$ permutes $U_{q+1}$,
where $\varepsilon_i \in U_{q+1}$
and $t_i \ge \deg(M_i)$. Let 
$H = M_1^{n t_2} \big( M_2 \circ x^n \circ L_1/M_1 \big)$ and
\[
  f(x) = x^r H(x^{q-1})^{m_1},
\]
where $n$, $r \in \n$, $m_1 = (r, q-1)$,  
$r/m_1 \equiv n t_1 t_2 \pmod{q+1}$. 
Then $f$ is \mfqtwostar \ifa $m_1 \mid m$ and 
$(n, q+1) = m / m_1$, where $1 \le m \le m_1 (q+1)$.
\end{theorem}

\begin{proof}
Since $M_1$ and $M_2$ have no roots in $U_{q+1}$ 
and $L_1/M_1$ permutes $U_{q+1}$, it follows that
$H$ has no roots in $U_{q+1}$.
Let $M_2 = \textstyle{\sum}\, a_j x^j \in \fqtwox$. Then
\begin{align*}
    H & = M_1^{n t_2} \big(\textstyle{\sum}\, 
        a_j x^j \circ x^n \circ L_1/M_1 \big) \\
    & = M_1^{n t_2} \textstyle{\sum}\, a_j (L_1/M_1)^{n j} \\ 
    & = \textstyle{\sum}\, a_j L_1^{n j} M_1^{n(t_2-j)}.
\end{align*}
For $x \in U_{q+1}$, $L_i = \varepsilon_i x^{t_i} M_i^q$ 
implies that
$L_1^q = \varepsilon_1^{-1} x^{-t_1} M_1$ and 
$\varepsilon_{2}^{-1} L_2 = x^{t_2} M_2^q 
  = \textstyle{\sum}\, a_j^q x^{t_2-j}$. Thus,
\begin{align*}
    H^q 
    & = \textstyle{\sum}\, a_j^q (L_1^q)^{nj} (M_1^q)^{n(t_2-j)} \\
    & = \textstyle{\sum}\, a_j^q (\varepsilon_1^{-1} x^{-t_1} M_1)^{n j} 
        (\varepsilon_1^{-1} x^{-t_1} L_1)^{n(t_2-j)}\\ 
    & = (\varepsilon_1^{-1} x^{-t_1})^{n t_2} \textstyle{\sum}\,
         a_j^q M_1^{n j} L_1^{n(t_2-j)} \\
    & = (\varepsilon_1^{-1} x^{-t_1})^{n t_2} M_1^{n t_2} 
        \textstyle{\sum}\, a_j^q (L_1/M_1)^{n(t_2-j)} \\
    & = \varepsilon_1^{-n t_2} x^{-n t_1 t_2} M_1^{n t_2} 
        \big(\textstyle{\sum}\, 
        a_j^q x^{t_2-j} \circ (L_1/M_1)^{n} \big) \\ 
    & = \varepsilon_1^{-n t_2} x^{-n t_1 t_2} M_1^{n t_2} 
        \big(\varepsilon_{2}^{-1} L_2 \circ L_1^n / M_1^n\big) \\ 
    & =  x^{-n t_1 t_2} M_1^{n t_2} 
        \big(\beta L_2 \circ L_1^n / M_1^n\big), 
\end{align*}
where $\beta = \varepsilon_1^{-n t_2} \varepsilon_2^{-1}$. 
For $x \in U_{q+1}$, $x^{r/m_1} = x^{n t_1 t_2}$ 
by $r/m_1 \equiv n t_1 t_2 \pmod{q+1}$, and so
\begin{align*}
    g(x) & \coloneqq x^{r/m_1} H^q/H  
     = \frac{\beta L_2 \circ L_1^n / M_1^n}{M_2 \circ L_1^n / M_1^n} \\ 
    & = \beta L_2/M_2 \circ x^n \circ L_1/M_1.
\end{align*}
Since $\beta L_2/M_2$ permutes $U_{q+1}$, we get 
\[
 (\beta L_2/M_2)^{-1} \circ g
 = x^n \circ L_1/M_1.
\]
Note that $f(x) \in U_{\frac{q^2-1}{m_1}}$ 
for any $x \in \fqtwostar$
and $g(x) \in U_{q+1}$ for $x \in U_{q+1}$. 
Thus the following diagrams are commutative: 
\[
  \xymatrix{
  \fqtwostar \ar[rr]^{f}\ar[d]_{x^{q-1}}  &   &  
  U_{\frac{q^2-1}{m_1}} \ar[d]^{x^{\frac{q-1}{m_1}}} \\
  U_{q+1} \ar[rr]^{g} \ar[d]_{L_1/M_1}  &   &  U_{q+1}
  \ar[d]^{(\beta L_2/M_2)^{-1}}  \\
  U_{q+1} \ar[rr]^{x^n}            &   &  U_{q+1}.
  }
\] 
Let $\lambda = L_1/M_1$ and 
$\bar{\lambda} = (\beta L_2/M_2)^{-1}$.
Since both $\lambda$ and $\bar{\lambda}$ permute $U_{q+1}$, 
$\# \lambda^{-1}(\alpha) = 
\# \bar{\lambda}^{-1}(\alpha^n)$ 
and $g$ is \mfield{1}{\lambda^{-1}(\alpha)} 
for any $\alpha \in U_{q+1}$.
By \cref{xrh(xs)2}, $f$ is \mfqtwostar \ifa 
$m_1 \mid m$, $(q-1)((q+1) \bmod{m_2}) < m$, 
and $x^n$ is \mbox{$m_2$-to-$1$} on $U_{q+1}$,
or equivalently $m_1 \mid m$ and $(n, q+1) = m_2$, 
where $1 \le m \le m_1(q+1)$ and $m_2 = m /m_1$.
\end{proof}

The conditions in \cref{xrhxs(m1)_Uq+1} can be satisfied.
Indeed, all the desired polynomials $L_i$ and $M_i$ 
are completely determined in 
{\cite[Lemma~2.1]{Zieve130776}}  
and {\cite[Proposition~3.5]{Bartoli18LqM}}
when $\deg(L_i) = \deg(M_i) = t_i \in \{1, 2\}$.
The next result is a reformulation of 
{\cite[Lemma~2.1]{Zieve130776}}.

\begin{lemma}\label{lem:RFUq+1toUq+1}
Let $\ell(x) \in \overline{\mathbb{F}}_q (x)$ 
be a degree-one rational function,
where $\overline{\mathbb{F}}_q$ 
is the algebraic closure of $\fq$. 
Then $\ell(x)$ permutes $U_{q+1}$ \ifa 
$\ell(x) = (\beta^q x + \alpha^q)/(\alpha x + \beta)$, 
where $\alpha$, $\beta \in \fqtwo$ and 
$\alpha^{q+1} \ne \beta^{q+1}$.
\end{lemma}

\cref{xrhxs(m1)_Uq+1} 
reduces to the following form 
when $L_2 = \beta^q x +  \alpha^q$ 
and $M_2 = \alpha x + \beta$.

\begin{corollary}\label{xrhxs(m1)_Uq+1_R1}
Let $L$, $M \in \fqtwox$ satisfy that $M$ has no roots in $U_{q+1}$,
$L = \varepsilon x^{t} M^q$ for any $x \in U_{q+1}$,  
and $L/M$ permutes $U_{q+1}$,
where $\varepsilon \in U_{q+1}$ and $t \ge \deg(M)$. Let
\[
  H = \alpha L^n + \beta M^n \qtq{and}
  f(x) = x^r H(x^{q-1})^{m_1},
\]
where $n \ge 1$, $\alpha$, $\beta \in \fqtwo$ with 
$\alpha^{q+1} \ne \beta^{q+1}$, $m_1 = (r, q-1)$, 
and $r / m_1 \equiv n t \pmod{q+1}$.
Then $f$ is \mfqtwostar \ifa
$m_1 \mid m$ and $(n, q+1) = m / m_1$,
where $1 \le m \le m_1 (q+1)$.
\end{corollary}

\begin{proof}
Take $M_2 = \alpha x + \beta$, 
$\varepsilon_2 = t_2 = 1$,  
and $L_2 = \beta^q x +  \alpha^q$.
Then $M_2$ has no roots in $U_{q+1}$ 
by $\alpha^{q+1} \ne \beta^{q+1}$,
$L_2 / M_2$ permutes $U_{q+1}$ 
by \cref{lem:RFUq+1toUq+1}, and 
$H = M_1^n ( M_2 \circ L_1^n / M_1^n ) 
= \alpha L_1^n + \beta M_1^n$. 
Then the result follows from 
\cref{xrhxs(m1)_Uq+1}.
\end{proof}

\begin{remark}
  In the case $\deg(L) = \deg(M) = t$ and $m_1 = m = 1$, \cref{xrhxs(m1)_Uq+1_R1} is equivalent to \cite[Theorem~3.3]{Bartoli18LqM}.
  In other cases, it generalizes 
  \cite[Theorem~3.3]{Bartoli18LqM}.
  Moreover, the proof of \cite[Theorem~3.3]{Bartoli18LqM} 
  mainly takes advantage of some properties of 
  ``$\beta$-associated polynomials", while \cref{xrhxs(m1)_Uq+1_R1} 
  is based on the commutative diagrams in the proof of 
  \cref{xrhxs(m1)_Uq+1}.
\end{remark}

In \cref{xrhxs(m1)_Uq+1_R1}, take $M = \gamma x + \delta$, 
$\varepsilon = t = 1$, and $L = \delta^q x +  \gamma^q$,
where $\gamma^{q+1} \ne \delta^{q+1}$.
Then $L/M$ permutes $U_{q+1}$ by \cref{lem:RFUq+1toUq+1}, 
and so we obtain the next result.

\begin{example}\label{xrhxs(m1)_Uq+1_R1L1}
Let $\alpha$, $\beta$, $\gamma$, $\delta \in \fqtwo$ 
satisfy $\alpha^{q+1} \ne \beta^{q+1}$ and 
$\gamma^{q+1} \ne \delta^{q+1}$. 
Let $f(x) = x^r H(x^{q-1})^{m_1}$, where 
\[
  H(x) = \alpha (\delta^q x +  \gamma^q)^n 
     + \beta (\gamma x + \delta)^n,
\]
$n, r \ge 1$,  $m_1 = (r, q-1)$, 
and $r / m_1 \equiv n \pmod{q+1}$.
Then~$f$ is \mfqtwostar \ifa 
$m_1 \mid m$ and $(n, q+1) = m / m_1$,
where $1 \le m \le m_1 (q+1)$.
\end{example} 

\begin{remark}
In the special case $\alpha \beta \gamma \delta \ne 0$
and $m_1 = m = 1$, 
\cref{xrhxs(m1)_Uq+1_R1L1} is equivalent to \cite[Theorem~1.2]{DingZ23}.
See related results in \cite{FuFLW19,PanarioUW23} 
and their generalization \cite{FengWang26}.
\end{remark}

In \cref{xrhxs(m1)_Uq+1_R1}, 
take $M = x^4 + x + 1$, $\varepsilon = 1$, 
$t = 5$, and $L = x^5 + x^4 + x$.
If $q = 2^s$ with $s \not\equiv 2 \pmod 4$, 
then $L/M$ permutes $U_{q+1}$ by \cref{g5to1}, 
and so we have the following result. 

\begin{example}\label{xrhxs(m1)_Uq+1_R1L5}
Let $q = 2^s$ with $s \not\equiv 2 \pmod 4$ 
and $\alpha$, $\beta \in \fqtwo$ with 
$\alpha^{q+1} \ne \beta^{q+1}$. 
Let $f(x) = x^r H(x^{q-1})^{m_1}$, where
\[
  H(x) = \alpha (x^5 + x^4 + x)^n 
        + \beta (x^4 + x + 1)^n,
\]
$n \ge 1$, $m_1 = (r, q-1)$, 
and $r / m_1 \equiv 5n \pmod{q+1}$.
Then~$f$ is \mfqtwostar \ifa
$m_1 \mid m$ and $(n, q+1) = m / m_1$,
where $1 \le m \le m_1 (q+1)$.
\end{example}

\cref{xrhxs(m1)_Uq+1} reduces to the next result
when $L_2 = c x^3 + x^2 + 1$ and $M_2 = x^3 + x + c$.

\begin{corollary}\label{xrhxs(m1)_Uq+1_R3}
Let $q$ be even and $L$, $M \in \fqtwox$ satisfy that 
$M$ has no roots in $U_{q+1}$,
$L = \varepsilon x^{t} M^q$ for any $x \in U_{q+1}$,  
and $L/M$ permutes $U_{q+1}$,
where $\varepsilon \in U_{q+1}$ and $t \ge \deg(M)$. Let
\[
  H = L^{3n} + L^{n} M^{2n} + c M^{3n}
  ~\text{and}~
  f(x) = x^r H(x^{q-1})^{m_1},
\]
where $n \ge 1$, 
$c \in \fqstar$ with $\tr_{q/2}(1 + c^{-1}) = 0$,
$m_1 = (r, q-1)$, and $r/m_1 \equiv 3 n t \pmod{q+1}$.
Then $f$ is \mfqtwostar \ifa $m_1 \mid m$ 
and $(n, q+1) = m/m_1$, where $1 \le m \le m_1(q+1)$.
\end{corollary}
\begin{proof}
Take $M_2 = x^3 + x + c$, $\varepsilon_2 = 1$, 
$t_2 = 3$, and $L_2 = c x^3 + x^2 + 1$.
Then $L_2 / M_2$ permutes $U_{q+1}$ by \cref{g3to1}  
and $H = M_1^{3n} ( M_2 \circ L_1^n / M_1^n ) 
= L_1^{3n} + L_1^{n} M_1^{2n} + c M_1^{3n}$. 
Now the result follows from \cref{xrhxs(m1)_Uq+1}.
\end{proof}

In \cref{xrhxs(m1)_Uq+1_R3}, 
taking $L = \beta^q x + \alpha^q$ 
and $M = \alpha x + \beta$ 
yields the next result.

\begin{example}\label{xrhxs(m1)_Uq+1_R3L1} 
Let $q$ be even and $\alpha$, $\beta \in \fqtwo$  
with $\alpha^{q+1} \ne \beta^{q+1}$. 
Let $f(x) = x^r H(x^{q-1})^{m_1}$, where
\[
  H(x) = (\beta^q x +  \alpha^q)^{3n} 
       + (\beta^q x +  \alpha^q)^{n} 
         (\alpha x + \beta)^{2n} 
       + c(\alpha x + \beta)^{3n},  
\]
$n \ge 1$, $c \in \fqstar$ with $\tr_{q/2}(1 + c^{-1}) = 0$,
$m_1 = (r, q-1)$, and $r/m_1 \equiv 3 n \pmod{q+1}$.
Then $f$ is \mfqtwostar \ifa $m_1 \mid m$ 
and $(n, q+1) = m/m_1$, where $1 \le m \le m_1(q+1)$.
\end{example} 

\subsection{\texorpdfstring{$\lambda$ is \onetoone from $U_{q+1}$ to $\fq \cup \{\infty\}$}%
{lambda is 1-to-1 from Uq+1 to Fq U infinity}}

In order to create a well-defined setting,
we stipulate that $c / 0 =  \infty$ 
for any $c \in \fqtwostar$. 
For any nonconstant $f = L / M$, 
where $L$, $M \in \fqtwox \setminus \{0\}$, define
\[
f(\infty) = 
\begin{cases}
\infty & \text{if $\deg(L) > \deg(M)$,} \\
a / b  & \text{if $\deg(L) = \deg(M)$,}  \\
0      & \text{if $\deg(L) < \deg(M)$,}
\end{cases}
\]
where $a$ and $b$ are the leading coefficients 
of $L$ and $M$, respectively.
In particular, $\infty^n = \infty$ for any $n \in \n$.
For arbitrary $N(x) = \sum_{i=0}^{u} 
a_i x^i \in \fqtwox$, define 
$N^{(q)}(x) = \sum_{i=0}^{u} a_i^q x^i$.

\begin{theorem}\label{xrhxs(m1)_Fq} 
Let $L$, $M \in \fqtwox$ satisfy that 
$L = \varepsilon x^{t} L^q$ and 
$M = \varepsilon x^{t} M^q$ 
for any $x \in U_{q+1}$ and that
$L/M$ induces a bijection from 
$U_{q+1}$ to $\fq \cup \{\infty\}$, 
where $\varepsilon \in U_{q+1}$ 
and $t \ge \max\{\deg(L), \deg(M)\}$.
Let $N \in \fqtwox$ satisfy that 
$N^{(q)} / N$ induces a bijection from 
$\fq \cup \{\infty\}$ to $U_{q+1}$. Let 
\[
  H = M^{n u} \big( N \circ x^n \circ L/M \big)
  \qtq{and}
  f(x) = x^r H(x^{q-1})^{m_1},
\]
where $n, r \in \n$, $u = \deg(N)$, 
$H$ has no roots in $U_{q+1}$,
$m_1 = (r, q-1)$, and $r/m_1 \equiv n t u \pmod{q+1}$. 
Then for $1 \le m \le m_1 (q+1)$,
$f$ is \mfqtwostar \ifa one of the following holds:
\begin{enumerate}[\upshape(1)]
\item $m = m_1$ and $(n, q-1) = 1$;
\item $m_1 \mid m$, $(n, q-1) = m / m_1 \ge 3$, and $2(q-1) < m$.
\end{enumerate}
\end{theorem}
\begin{proof}
Put $N = \sum_{i=0}^{u} a_i x^i \in \fqtwox$. Then
\[
  H = M^{n u} (\sum a_i x^i \circ x^n \circ L/M )
    = \sum a_i L^{n i} M^{n (u-i)} 
\]
and for any $x \in U_{q+1}$, 
\begin{align*}
  H^q 
  & = \sum a_i^q L^{q n i} M^{q n (u-i)} \\
  & = \sum a_i^q 
      (\varepsilon^{-1} x^{-t} L)^{n i} 
      (\varepsilon^{-1} x^{-t} M)^{n (u-i)} \\
  & = (\varepsilon^{-1} x^{-t})^{n u} 
      \sum a_i^q L^{n i} M^{n (u-i)}.
\end{align*}
Define $g(x) = x^{r/m_1} H^{q-1}$.
The condition $r/m_1 \equiv n t u \pmod{q+1}$ 
implies $x^{r/m_1} = x^{n t u}$ 
for any $x \in U_{q+1}$.
Recall that $H$ has no roots in $U_{q+1}$. 
Thus, for any $x \in U_{q+1}$,
\begin{equation}\label{g=sumaqa}
g(x) = x^{r/m_1} H^q / H  
= \frac{\beta \sum_{i=0}^{u} a_i^q L^{n i} M^{n (u-i)}}
            {~\sum_{i=0}^{u} a_i L^{n i} M^{n (u-i)}},
\end{equation}
where $\beta = \varepsilon^{-n u}$.      
If $M(x) \ne 0$ for some $x \in U_{q+1}$, then 
\begin{equation}\label{g=LxnR}
\begin{aligned}
  g(x)  
  & = \frac{\beta \sum a_i^q (L/M)^{n i} }
         {~\sum a_i (L/M)^{n i} } 
   = \frac{\beta N^{(q)} \circ (L/M)^{n}}
                       {N \circ (L/M)^{n}} \\ 
  & = \frac{\beta N^{(q)}}{N} \circ x^n \circ \frac{L}{M}.
\end{aligned}
\end{equation}
If $M(x_0) = 0$ for some $x_0 \in U_{q+1}$, 
then $x_0$ is unique and $L(x_0) \ne 0$,
since $L/M$ induces a bijection from 
$U_{q+1}$ to $\fq \cup \{\infty\}$. 
Hence by \cref{g=sumaqa},
\[
g(x_0) = \beta a_u^q L(x_0)^{n u} / a_u L(x_0)^{n u} 
= \beta a_u^q / a_u.
\]
Because $L(x_0) \ne 0$ and $M(x_0) = 0$, we get 
$L(x_0) / M(x_0) = \infty$ and $\infty^n = \infty$. Thus, 
\[
\frac{\beta N^{(q)}}{N} \circ x^n \circ \frac{L(x_0)}{M(x_0)}
= \frac{\beta N^{(q)}(\infty)}{N(\infty)} 
= \frac{\beta a_u^q}{a_u}.
\]
In summary, \cref{g=LxnR} holds for any $x \in U_{q+1}$.
Since $\beta N^{(q)} / N$ induces a bijection 
from $\fq \cup \{\infty\}$ to $U_{q+1}$, 
\[
 (\beta N^{(q)} / N)^{-1} \circ g
 = x^n \circ L/M.
\]
Note that $f(x) \in U_{\frac{q^2-1}{m_1}}$ 
for $x \in \fqtwostar$
and $g(x) \in U_{q+1}$ for $x \in U_{q+1}$. 
Thus the following diagrams are commutative: 
\[
  \xymatrix{
  \fqtwostar \ar[rr]^{f}\ar[d]_{x^{q-1}}  &   &  
  U_{\frac{q^2-1}{m_1}} \ar[d]^{x^{\frac{q-1}{m_1}}} \\
  U_{q+1} \ar[rr]^{g} \ar[d]_{L/M}  &   &  U_{q+1}
  \ar[d]^{(\beta N^{(q)}/N)^{-1}}  \\
  \fq \cup \{\infty\} \ar[rr]^{x^n}  &   &  \fq \cup \{\infty\},
  }
\] 
Let $\lambda = L/M$ and 
$\bar{\lambda} = (\beta N^{(q)}/N)^{-1}$.
Since both $\lambda$ and $\bar{\lambda}$ are bijective, 
$\# \lambda^{-1}(\alpha) = \# \bar{\lambda}^{-1}(\alpha^n) = 1$ 
and $g$ is \mfield{1}{\lambda^{-1}(\alpha)} 
for any $\alpha \in \fq \cup \{\infty\}$.
By \cref{xrh(xs)2}, for $1 \le m \le m_1 (q+1)$,
$f$ is \mfqtwostar \ifa 
$m_1 \mid m$, $(q-1)((q+1) \bmod{m_2}) < m$, 
$x^n$ is \mfield{m_2}{\fq \cup \{\infty\}}, and 
$\# E_{x^n}(\fq \cup \{\infty\}) = (q+1) \bmod{m_2}$,
where $m_2 = m /m_1$.

Under the condition $m_1 \mid m$, if $m_2 = 1$, then
$f$ is \mfqtwostar \ifa $(n, q-1) = 1$.
If $m_2 = 2$, then $x^n$ only maps 0 to 0 and $\infty$ to $\infty$.
Hence $x^n$ is not \mfield{2}{\fq \cup \{\infty\}}, 
and so~$f$ is not \mfqtwostar.
If $m_2 \ge 3$, then $f$ is \mfqtwostar \ifa
$(q-1)((q+1) \bmod{m_2}) < m$ and $(n, q-1) = m_2$, i.e., 
$(n, q-1) = m_2$ and $2(q-1) < m$.
\end{proof}

\begin{remark}
The idea of \cref{xrhxs(m1)_Fq} comes from~\cite{Bartoli18LqM}.
In the case $t = \deg(L) = \deg(M)$ and $m_1 = m = 1$, 
\cref{xrhxs(m1)_Fq} is similar to \cite[Theorem~5.1]{Bartoli18LqM}.
In other cases, \cref{xrhxs(m1)_Fq} generalizes 
\cite[Theorem~5.1]{Bartoli18LqM}.
\end{remark}

All degree-one rational functions over $\fqtwo$ that
are bijections from $U_{q+1}$ to $\fq \cup \{\infty\}$ 
are completely determined in 
{\cite[Lemma~3.1]{Zieve130776}},
which can be reformulated as follows.

\begin{lemma}\label{lem:RFUq+1toinfty}
Let $\ell(x) \in \overline{\mathbb{F}}_q (x)$ 
be a degree-one rational function,
where $\overline{\mathbb{F}}_q$ is the algebraic closure of $\fq$. 
Then $\ell(x)$ induces a bijection from $U_{q+1}$ to 
$\fq \cup \{\infty\}$ \ifa 
\[
\ell(x) = (\beta x +  \beta^q)/(\alpha x + \alpha^q),
\] 
where $\alpha$, $\beta \in \fqtwostar$ 
and $\alpha^{q-1} \ne \beta^{q-1}$.
\end{lemma}

\cref{xrhxs(m1)_Fq}
reduces to the following form 
if $N = \alpha x + \beta$.

\begin{corollary}\label{xrhxs(m1)_Fq_R1}
Let $L$, $M \in \fqtwox$ satisfy that 
$L = \varepsilon x^{t} L^q$ and 
$M = \varepsilon x^{t} M^q$ for any $x \in U_{q+1}$,
$L/M$ induces a bijection from 
$U_{q+1}$ to $\fq \cup \{\infty\}$, 
where $\varepsilon \in U_{q+1}$ 
and $t \ge \max\{\deg(L), \deg(M)\}$. Let 
\[
  H = \alpha L^n + \beta M^n 
  \qtq{and}
  f(x) = x^r H(x^{q-1})^{m_1},
\]
where $n \ge 1$, $\alpha$, $\beta \in \fqtwostar$ 
with $\alpha^{q-1} \ne \beta^{q-1}$,
$H$ has no roots in $U_{q+1}$,
$m_1 = (r, q-1)$, and $r/m_1 \equiv n t \pmod{q+1}$. 
Then
$f$ is \mfqtwostar \ifa $m = m_1$ and $(n, q-1) = 1$, 
where 
$1 \le m \le \min \{2(q-1), m_1 (q+1)\}$.
\end{corollary}
\begin{proof}
Let $\ell(x) = (\beta x +  \beta^q)/(-\alpha x -\alpha^q)$.
Then it induces a bijection from $U_{q+1}$ to $\fq \cup \{\infty\}$
by \cref{lem:RFUq+1toinfty}, and its compositional inverse is 
$\ell^{-1}(x) = -(\alpha^q x + \beta^q)/(\alpha x + \beta)$,
which induces a bijection from $\fq \cup \{\infty\}$ to $U_{q+1}$.
In \cref{xrhxs(m1)_Fq}, take $N = \alpha x + \beta$.
Then $N^{(q)} = \alpha^q x + \beta^q$ 
and $H = M^n ( N \circ L^n / M^n ) 
= \alpha L^n + \beta M^n$. 
Now the result follows from \cref{xrhxs(m1)_Fq}.
\end{proof}

Substituting the rational function 
in \cref{lem:RFUq+1toinfty} 
to \cref{xrhxs(m1)_Fq_R1} 
yields the next result.

\begin{example}\label{xrhxs(m1)_Fq_R1L1}
Let $\alpha$, $\beta$, $\gamma$, $\delta \in \fqtwostar$ 
satisfy $\alpha^{q-1} \ne \beta^{q-1}$ and 
$\gamma^{q-1} \ne \delta^{q-1}$. Let 
$f(x) = x^r H(x^{q-1})^{m_1}$, where
\[
  H(x) = \alpha (\gamma x + \gamma^q)^n 
        + \beta (\delta x + \delta^q)^n,
\]
$n$, $r \ge 1$, $m_1 = (r, q-1)$, 
and $r/m_1 \equiv n \pmod{q+1}$. 
Then~$f$ is \mfqtwostar \ifa $m = m_1$ 
and $(n, q-1) = 1$, where 
$1 \le m \le \min \{2(q-1), m_1 (q+1)\}$.
\end{example}
\begin{proof}
Take $L = \gamma x + \gamma^q$ 
and  $M = \delta x + \delta^q$.
Then $L/M$ induces a bijection from $U_{q+1}$ to 
$\fq \cup \{\infty\}$ by \cref{lem:RFUq+1toinfty},
and $\alpha L(x)^n + \beta M(x)^n$ has no roots in $U_{q+1}$.
Indeed, if $\alpha L(x_0)^n = -\beta M(x_0)^n$
for some $x_0 \in U_{q+1}$,
then $L(x_0) \ne 0$ and $M(x_0) \ne 0$ by 
$\alpha \beta \ne 0$ and $\gamma^{q-1} \ne \delta^{q-1}$.
Thus, $-\beta/\alpha = (L(x_0)/M(x_0))^n \in \fqstar$,
contrary to $\alpha^{q-1} \ne \beta^{q-1}$. 
Then the result follows from \cref{xrhxs(m1)_Fq_R1}.
\end{proof}

\begin{remark}
In the case $m_1 = m = 1$, \cref{xrhxs(m1)_Fq_R1L1} 
is equivalent to \cite[Theorem~1.1]{DingZ23}. 
In other cases, \cref{xrhxs(m1)_Fq_R1L1} 
is an extension of some results in \cite{DingZ23,FuFLW19,PanarioUW23}.
\end{remark}

\begin{remark}\label{rem:g=hx3h}
We now give different proofs of \cref{g3to1,g5to1}
by using the composition of rational functions. 
Let $q = 2^n$ and let $a \in \fqtwo$ 
be a solution of $x^2 + c x + 1 = 0$ and 
$\lambda(x) = (x + a)/(a x + 1)$. Then the 
compositional inverse of $\lambda$ is itself. 
By $a^2 = a c + 1$ and $a^3 = a^2 c + a$, 
it is easy to verify that 
$\lambda \circ g = x^3 \circ \lambda$, i.e.,
$g = \lambda \circ x^3 \circ \lambda$,
where $g$ is as in \cref{g3to1}.
For any $x \in U_{q+1}$, $g(x)^q = g(x)^{-1}$
and so $g$ maps $U_{q+1}$ to itself.
Hence, the following diagram is commutative:
\[
\xymatrix{
  U_{q+1} \ar[rr]^{g}\ar[d]_{\lambda}  &   &  
  U_{q+1} \ar[d]^{\lambda} \\
  \lambda(U_{q+1}) \ar[rr]^{x^3}   &   &  
  \lambda(U_{q+1}).}
\]
If $\tr_{q/2}(1/c) = 0$, then 
$a \in \fq \setminus \{0, 1\}$ by 
\cref{tr(1/c)}, and so $a^{q+1} = a^2 \ne 1$. 
By \cref{lem:RFUq+1toUq+1}, $\lambda$ permutes 
$U_{q+1}$ and so $\lambda(U_{q+1}) = U_{q+1}$.
When $n$ is odd, $(3, q+1) = 3$ and so $x^3$ is
\mset{3}{U_{q+1}}. Thus, $g$ is \mset{3}{U_{q+1}}.
When $n$ is even, $(3, q+1) = 1$ and so $x^3$ is
\mset{1}{U_{q+1}}. Thus, $g$ is \mset{1}{U_{q+1}}.
If $\tr_{q/2}(1/c) = 1$, then 
$a \in U_{q+1} \setminus \{1\}$ by \cref{tr(1/c)},
and so $a = e^{q-1}$ for some $e \in \fqtwostar$. Then
\[
\lambda(x) = \frac{e x + e a}{e a x + e} 
= \frac{e x + e^q}{e^q x + e}
\]
and $e^{q(q-1)} \ne e^{q-1}$. 
Then by \cref{lem:RFUq+1toinfty}, $\lambda$ induces 
a bijection from $U_{q+1}$ onto $\fq \cup \{\infty\}$,
and so $\lambda(U_{q+1}) = \fq \cup \{\infty\}$.
When $n$ is odd, $(3, q-1) = 1$ and so $x^3$ permutes  
$\fq \cup \{\infty\}$. Thus, $g$ is \mset{1}{U_{q+1}}.
When $n$ is even, $(3, q-1) = 3$ and so $x^3$ is
\thrtoone form $\fq \cup \{\infty\}$ from to itself. 
Thus, $g$ is \mset{3}{U_{q+1}}.
This completes the proof of \cref{g3to1}. 
In the above analysis, 
taking $c=1$ and replacing~$3$ by~$5$ 
yield another proof of \cref{g5to1}.
\end{remark}

We next construct a class of rational functions 
from $\fq \cup \{\infty\}$ to $U_{q+1}$ by the 
composition of monomials and degree-one rational functions.
Take $\alpha$, $\beta \in \fqtwostar$ with 
$\alpha^{q-1} \ne \beta^{q-1}$~and 
\[
\ell_2(x) = -x \circ 
    \frac{\beta x + \beta^q}{\alpha x + \alpha^q} \circ -x 
= \frac{\beta x - \beta^q}{-\alpha x + \alpha^q}.
\]
By \cref{lem:RFUq+1toinfty}, $\ell_2$ induces a bijection 
from $U_{q+1}$ to $\fq \cup \{\infty\}$.
Then its compositional inverse is
\[
\ell_2^{-1} (x) = (\alpha^q x + \beta^q)/(\alpha x + \beta),
\]
which induces a bijection from $\fq \cup \{\infty\}$ to $U_{q+1}$. 
Let $k \in \n$ and $(k, q+1) = 1$. Then $x^k$ permutes $U_{q+1}$. Pick 
\[
\ell_1(x) = (\gamma^q x + \delta^q)/(\delta x + \gamma),
\]
where $\gamma, \delta \in \fqtwo$ with 
$\gamma^{q+1} \ne \delta^{q+1}$.
Then $\ell_1$ permutes $U_{q+1}$ 
by \cref{lem:RFUq+1toUq+1}. Let 
\begin{align*}
  \lambda_k(x) 
  & = \ell_1 \circ x^k \circ \ell_2^{-1} 
    = \frac{\gamma^q(\alpha^q x + \beta^q)^k 
        + \delta^q(\alpha x + \beta)^k}
      {\gamma(\alpha x + \beta)^k 
      + \delta(\alpha^q x + \beta^q)^k}, 
\end{align*}
i.e., the following diagram is commutative: 
\[
  \xymatrix{
  \fq \cup \{\infty\} \ar[rr]^{\lambda_k} \ar[d]_{\ell_2^{-1}}  
  &   &  U_{q+1}  \\
  U_{q+1} \ar[rr]^{x^k}  &   &  U_{q+1} \ar[u]_{\,\ell_1}.
  }
\]
Then $\lambda_k$ induces a bijection 
from $\fq \cup \{\infty\}$ to $U_{q+1}$. 
Applying \cref{xrhxs(m1)_Fq} to 
$N = \gamma(\alpha x + \beta)^k 
+ \delta(\alpha^q x + \beta^q)^k$ 
yields the following result.

\begin{corollary}\label{xrhxs(m1)_Fq_Rk}
Let $L$, $M \in \fqtwox$ satisfy that 
$L = \varepsilon x^{t} L^q$ and 
$M = \varepsilon x^{t} M^q$ for any $x \in U_{q+1}$,
$L/M$ induces a bijection from 
$U_{q+1}$ to $\fq \cup \{\infty\}$, 
where $\varepsilon \in U_{q+1}$ 
and $t \ge \max\{\deg(L), \deg(M)\}$. 
Assume $\alpha$, $\beta \in \fqtwostar$, $\gamma$, 
$\delta \in \fqtwo$, $\alpha^{q-1} \ne \beta^{q-1}$, 
and $\gamma^{q+1} \ne \delta^{q+1}$. Let 
$f(x) = x^r H(x^{q-1})^{m_1}$, where
\[
  H = \gamma(\alpha L^n + \beta M^n)^k 
    + \delta(\alpha^q L^n + \beta^q M^n)^k,
\]
$n$, $k$, $r\ge 1$, $(k, q+1) = 1$,
$H$ has no roots in $U_{q+1}$,
$m_1 = (r, q-1)$, and $r/m_1 \equiv n t k\pmod{q+1}$. 
Then~$f$ is \mfqtwostar \ifa $m = m_1$ 
and $(n, q-1) = 1$, where 
$1 \le m \le \min \{2(q-1), m_1 (q+1)\}$.
\end{corollary}

Substituting the rational function 
in \cref{lem:RFUq+1toinfty} 
to \cref{xrhxs(m1)_Fq_Rk} 
yields the next result.

\begin{example}\label{xrhxs(m1)_Fq_RkL1}
Let $\beta$, $\theta \in \fqtwostar$ and 
$\delta \in \fqtwo$ satisfy $\beta^{q-1} \ne 1$, 
$\theta^{q-1} \ne 1$, and $\delta^{q+1} \ne 1$. 
Let $f(x) = x^r H(x^{q-1})^{m_1}$, where 
\[
  H(x) = 
   ( (x+1)^n + \beta   (\theta x + \theta^q)^n  )^k + \delta 
   ( (x+1)^n + \beta^q (\theta x + \theta^q)^n  )^k, 
\]
$n$, $k$, $r\ge 1$, $(k, q+1) = 1$, 
$m_1 = (r, q-1)$, and $r/m_1 \equiv kn \pmod{q+1}$. 
Then $f$ is \mfqtwostar \ifa $m = m_1$ and $(n, q-1) = 1$, 
where $1 \le m \le \min \{2(q-1), m_1 (q+1)\}$.
\end{example}

\begin{proof}
Let $L(x) = x+1$ and $M(x) = \theta x + \theta^q$.
By \cref{lem:RFUq+1toinfty}, $L/M$ induces a bijection 
from $U_{q+1}$ to $\fq \cup \{\infty\}$.
By the proof of \cref{xrhxs(m1)_Fq_R1L1},  
$L^n + \beta^q M^n$ has no roots in $U_{q+1}$. 
If $H(\bar{x}) = 0$ for some $\bar{x} \in U_{q+1}$, then
\begin{align*}
  - \delta 
  & = (L(\bar{x})^n + \beta M(\bar{x})^n)^k / 
      (L(\bar{x})^n + \beta^q M(\bar{x})^n)^k \\
  & = x^k \circ (x + \beta)/(x + \beta^q) 
      \circ x^n \circ L/M \circ \bar{x} \in U_{q+1}, 
\end{align*}
contrary to $\delta^{q+1} \ne 1$.
Thus $H$ has no roots in $U_{q+1}$.
Then the result follows from 
\cref{xrhxs(m1)_Fq_Rk}.
\end{proof}

There are many functions that permute 
$\fq \cup \{\infty\}$; see for example 
\cite{Hou15, WangIndex19, FengWang26, prfHou21deg4, 
ZhengMto1RFDeg3} and the references therein. 
By substituting these functions for~$x^n$ 
in \cref{xrhxs(m1)_Fq}, one can obtain more 
classes of \mtoone mappings over $\fqtwostar$.

\section{Recursive construction of \texorpdfstring{\mtoone}{m-to-1} mappings}

We first use \cref{constr3} to give a method
for constructing \mtoone mappings from other \mtoone mappings. 

\begin{theorem}\label{F=xmf}
Let $k$, $\ell$, $r$, $s$, $t \in \n$ satisfy 
$q-1 = \ell s$, $(r, s) \mid t$,
and $(r, s) = (r + k t, s)$.
Suppose $M \in \fqx$ satisfies
$\varepsilon x^{t/m_1} M(x)^{s/m_1} = 1$
for any $x \in U_{\ell}$, where $m_1 = (r, s)$
and $\varepsilon \in U_{\ell m_1}$. 
Let $f(x) = x^r h(x^s)$ and
\[
F(x) = x^{k t} M(x^s)^{k} f(x),
\]
where $h \in \fqx$ has no roots in $U_{\ell}$.
Then $F$ is \mfqstar \ifa $f$ is \mfqstar, 
where $1 \le m \le \ell m_1$.
\end{theorem}
\begin{proof}
Put $\lambda = x^s$, $\bar{\lambda} = x^{s_1}$,
and $g(x) = x^{r_1} h(x)^{s_1}$, 
where $r_1 = r /m_1$ and $s_1 = s/m_1$.
In the proof of \cref{MainThm}, we have shown that 
$\bar{\lambda} \circ f = g \circ \lambda$,
$\lambda$ is surjective from $\fqstar$ to $U_{\ell}$, 
$\# \lambda^{-1}(\alpha) 
= m_1 \, \# \bar{\lambda}^{-1}(g(\alpha))$,  
and~$f$ is \mfield{m_1}{\lambda^{-1}(\alpha)} 
for any $\alpha \in U_{\ell}$. 
For $x \in \lambda^{-1}(\alpha)$, 
i.e., $\lambda(x) = \alpha$, we obtain
$F(x) = x^{r + k t} M(\alpha)^{k} h(\alpha)$,
and thus $F$ is \mfield{m_1}{\lambda^{-1}(\alpha)}
since $(r + k t, s) = m_1$. 
Clearly, $\bar{\lambda}$ is a homomorphism 
from~$\fqstar$ onto~$U_{\ell m_1}$ and
\begin{align*}
(x^{k t} M_1^{k})^{s_1}  
 = (x^{s_1 t} M_1^{s_1})^{k}  
 = (x^{s t_1} M_1^{s_1})^{k}  
 = \varepsilon^{-k} \in U_{\ell m_1}
\end{align*}
for any $x \in \fqstar$, where $M_1 = M(x^s)$
and $t_1 = t/m_1$.
Then the result follows from \cref{constr3}.
\end{proof}

This result implies that if~$f$ is \mfqstar, then it 
can be used to construct a new \mtoone mapping~$F$.
In particular, applying \cref{F=xmf} to $\ell = q + 1$ 
and $s = q-1$ yields the next result.

\begin{theorem}\label{F=xmf_q2} 
Suppose $M \in \fqtwox$ has no roots in 
$U_{q+1}$ and $\varepsilon x^t M(x)^q = M(x)$
for any $x \in U_{q+1}$, 
where $\varepsilon \in U_{q+1}$ 
and $\deg(M) \le t \le 2 \deg(M)$.
Let $f(x) = x^r h(x^{q-1})$ and
\[
F(x) = x^{kt} M(x^{q-1})^{k} f(x),
\]
where $r$, $k \in \n$ satisfy 
$(r, q - 1) = (r+kt, q - 1) = 1$ 
and $h \in \fqtwox$ has no roots in $U_{q+1}$. 
Then $F$ is \mfqtwostar \ifa $f$ is \mfqtwostar,
where $1 \le m \le q+1$.
\end{theorem}

We next use \cref{F=xmf_q2,f3to1_h=x3+x+c} 
to construct new \thrtoone mappings.

\begin{corollary} \label{F3to1xt}
Let $f(x) = x^{3q} + x^{q+2} + c x^3$,
where $q = 2^n$ with odd $n \ge 3$ and $c \in \fqstar$.
Let 
\[
F(x) = x^{k t} (x^{t (q-1)} + a)^k f(x),
\]
where $k$, $t \in \n$ and $a \in U_{q+1} \setminus (U_{q+1})^{t}$.
Assume $(t, q+1) \ge 2$ and $(3 + k t, q-1) = 1$. 
Then $F$ is \mfield{3}{\fqtwostar} \ifa 
$\tr_{q/2}(1/c) = 0$.
\end{corollary} 

\begin{proof}
The condition $(t, q+1) \ge 2$ implies that
$(U_{q+1})^{t} \neq U_{q+1}$. 
Put $M(x) = x^t - a$. Since $a \in U_{q+1}$, 
we have $- a x^t M(x)^q = M(x)$ for any $x \in U_{q+1}$.
It follows from $a \notin (U_{q+1})^{t}$ 
that~$M$ has no roots in $U_{q+1}$.
Clearly $f(x) = x^3 h(x^{q-1})$, where 
$h(x) = x^3 + x + c$ has no roots in $U_{q+1}$.
Odd~$n$ implies that $(3, q-1) = 1$.
Then the result follows from 
\cref{F=xmf_q2,f3to1_h=x3+x+c}.
\end{proof}

The condition of $t$ and $k$ in \cref{F3to1xt}
can be met easily; for example, 
$n = 5$, $t = 3$ and $1 \le k \le 10$.

\begin{corollary} \label{F3to1hd}
Let $f(x) = x^{3q} + x^{q+2} + c x^3$,
where $q = 2^n$ with odd $n \ge 3$ and $c \in \fqstar$.
Assume 
\[
F(x) = x^{k(d-1)} h_d(x^{q-1})^k f(x),
\]
where $k \in \n$, $d$ is odd and $h_d$ is as in \cref{hd}.
Let $(d, q+1) = 1$ and $(3 + k(d-1), q-1) = 1$. 
Then $F$ is \mfield{3}{\fqtwostar} \ifa 
$\tr_{q/2}(1/c) = 0$.
\end{corollary} 

\begin{proof}
Since $(d, q+1) = 1$ and $d$ is odd, 
we get $(d, q(q+1)) = 1$ and so~$h_d$ 
has no roots in $U_{q+1}$ by \cref{lem:hdxeUL}.
Let $t = d-1$. Then $h_d(x) = x^t h_d(x)^q$ 
for any $x \in U_{q+1}$.
Because $n$ is odd, we have $(3, q-1) = 1$.
Then the result follows from 
\cref{F=xmf_q2,f3to1_h=x3+x+c}.
\end{proof}

The condition of $d$ and $k$ in \cref{F3to1hd}
can be met easily; for example, 
$n = 3$, $d = 5$ and $2 \le k \le 7$.

We remark that \cref{F=xmf,F=xmf_q2} can be used to construct  new \mtoone mappings recursively. Taking \cref{F=xmf_q2} as an example, let us start with any \mtoone mapping on $\fqtwostar$, say $f(x) = x^r h(x^{q-1})$ with $h \in \fqtwox$ has no roots in $U_{q+1}$. For any fixed $t$ and $M(x)$ satisfying the assumptions in \cref{F=xmf_q2}, there are many $k$'s such that $(r+kt, q-1)=1$. Pick any such $k$ and we obtain an \mtoone mapping on $\fqtwostar$,  i.e.,  $F(x) = x^{kt} M(x^{q-1})^{k} f(x)$. In fact,  
\[
F(x) =  x^{r+kt} M(x^{q-1})^kh(x^{q-1}).
\]
Because $(r+kt, q-1)=1$ and $M^k h$ has no roots in $U_{q+1}$, 
we can apply \cref{F=xmf_q2} again to pick~$t'$, $M'$, and $k'$ 
such that $(r + k t + k' t', q-1) = 1$ and then 
obtain another \mtoone mapping on $\fqtwostar$,
$F'(x) = x^{k' t'} M'(x^{q-1})^{k'} F(x)$.

We next give a method for constructing 
\mtoone mappings from \onetoone mappings.

\begin{theorem}
Let $q-1 = \ell s$ and let $M \in \fqx$ satisfy
$\varepsilon x^t M(x)^s = 1$ for any $x \in U_{\ell}$, 
where $\ell$, $s$, $t \in \n$ and $\varepsilon \in U_{\ell}$. 
Let $f(x) = x^r h(x^s)$ and $F(x) = x^{k t} M(x^s)^{k} f(x)$,
where $r$, $k \in \n$ and $h \in \fqx$. 
If~$f$ permutes $\fq$, 
then $F$ is \mfield{(r + k t, s)}{\fqstar}.
\end{theorem}
\begin{proof}
Clearly, $F(x) = x^{r + k t} M(x^s)^{k} h(x^s)$.
Put $m_1 = (r + k t, s)$ and 
$g(x) = x^{(r + k t)/m_1} (M(x)^k h(x))^{s/m_1}$.
Since $f$ permutes $\fq$ and $\varepsilon x^t M(x)^s = 1$ for any $x \in U_{\ell}$, 
it follows that $h$ and $M$ have no roots in $U_{\ell}$.
By \cref{xrh(xs):m2=1}, $F$ is \mfield{m_1}{\fqstar} 
\ifa $g$ is \mfield{1}{U_{\ell}}. 
For any $x \in U_{\ell}$,
\begin{align*}
  x^{m_1} \circ g(x) 
  & = x^{r + k t} M(x)^{k s} h(x)^s \\
  & = (x^{t} M(x)^{s})^k x^r h(x)^s \\
  & = \varepsilon^{-k} x^r h(x)^s.
\end{align*}
Since $f$ permutes $\fq$, we get $x^r h(x)^s$
permutes $U_{\ell}$ by \cref{xrhxs1to1}.
Hence $\varepsilon^{-k} x^r h(x)^s$ 
(i.e., $x^{m_1} \circ g(x)$) permutes $U_{\ell}$,
and thus $g$ is \mfield{1}{U_{\ell}}.
Therefore, $F$ is \mfield{m_1}{\fqstar}.
\end{proof}

This result establishes an important and interesting 
link between permutations and \mtoone mappings.
In other words, we can use known permutations of~$\fq$ 
to construct \mtoone mappings on $\fqstar$. 
We now give an example to illustrate this result
by employing a class of permutation binomials 
given by Hou~\cite{Hou32}.

\begin{corollary} \label{Fmto1xt}
Let $f(x) = x^{2q - 1} + c x \in \fqtwox$ 
with $q$ odd and $c^{(q+1)/2} = -1$ or~$3$.
Let $F(x) = x^{k t} (x^{t (q-1)} - a)^k f(x)$, 
where $k$, $t \in \n$, $(t, q+1) \ge 2$ 
and $a \in U_{q+1} \setminus (U_{q+1})^{t}$.
Then~$F$ is \mfield{m}{\fqtwostar}, 
where $m = (1 + k t, q-1)$.
\end{corollary} 

\begin{proof}
The fact that~$f$ permutes $\fqtwo$ 
is a part of \cite[Theorem~A]{Hou32}.
The remainder of the argument is similar to 
that in \cref{F3to1xt} and thus is omitted. 
\end{proof}

\section*{Conclusions}
  
In this paper, we introduced the definition of \mtoone mappings. We presented a generalized local criterion for a mapping to be \mtoone and proposed three constructions of \mtoone mappings, which unify and generalize all recent constructions in \cite{2to1-MesQ19,GAO21mto1,nto1-NiuLQL23,2to1-YuanZW21,Qin2469}. 
In particular, we characterized \mtoone mappings of the form $x^r h(x^{(q-1)/\ell})$ and established interesting links between \mtoone mappings on $\fqstar$ and $m_2$-to-$1$ mappings on $U_{\ell}$, where $m_2=m/m_1$ and $m_1 = (r, (q-1)/\ell)$. We also systematically studied many explicit classes of these \mtoone mappings for small $m$'s, or small index  $\ell$, or  special underlying mappings $g(x) = x^{r/m_1} h(x)^{(q-1)/\ell m_1}$ over $U_\ell$ such as monomials, rational functions, and those that can be further reduced through commutative diagrams. Finally, we provided two methods for constructing new 
\mtoone mappings from known permutations or \mtoone mappings.

\section*{Acknowledgment}

We deeply thank the anonymous reviewers 
and the Associate Editor Prof.\ Gohar M. Kyureghyan
for their valuable comments and suggestions 
which have greatly improved the manuscript.
We also thank Yang Zhang for his contribution to 
\cref{g3to1,g1_3to1,g5to1}.



\newpage
\begin{IEEEbiographynophoto}{Yanbin Zheng}
received the Ph.D. degree in Mathematics from 
Guangzhou University, Guangzhou, China, in 2015.
He is currently an Associate Professor with the School of 
Mathematical Sciences, Qufu Normal University, Qufu, China.
His research interests include finite fields and cryptography.
\end{IEEEbiographynophoto}

\vspace{-18pt}

\begin{IEEEbiographynophoto}{Yanjin Ding}
received the B.S. degree in mathematics from Jining University, Qufu, China, in 2020, and the M.S. degree in mathematics from Qufu Normal University, Qufu, China, in 2024.
She is currently pursuing the Ph.D. degree in mathematics with Hubei University, Wuhan, China. Her research interests include finite fields and cryptography.
\end{IEEEbiographynophoto}

\vspace{-18pt}

\begin{IEEEbiographynophoto}{Meiying Zhang}
received the B.S. and M.S. degrees in mathematics from
Qufu Normal University, Qufu, China, in 2020 and 2024, respectively. 
She is currently pursuing the Ph.D. degree in mathematics at Qufu Normal University.
Her research interests include finite fields and coding theory.
\end{IEEEbiographynophoto}

\vspace{-18pt}

\begin{IEEEbiographynophoto}{Pingzhi Yuan}
received the Ph.D. degree in mathematics from Sichuan University, Chengdu, China, in 1997. 
He is currently a Professor with the School of Mathematical Sciences, South China Normal University, Guangzhou, China. 
His research interests include finite fields and zero-sum theory.
\end{IEEEbiographynophoto}

\vspace{-18pt}

\begin{IEEEbiographynophoto}{Qiang Wang}
received the B.S. and M.S. degrees in mathematics from 
Shaanxi Normal University, Xi'an, China, in 1993 and 1996 respectively, 
the M.S. degree in information and system science from 
Carleton University, Ottawa, Canada, in 2002,
and the Ph.D. degree in mathematics from the Memorial University of
Newfoundland, Canada, in 2000. 
He is currently a Professor at Carleton University.
His main research interests are finite fields and applications 
in coding theory, combinatorics, and cryptography. 
He serves as an Associate Editor for journals including
\emph{Applicable Algebra in Engineering, Communication and Computing} (Springer),   
\emph{Finite Fields and Their Applications} (Elsevier), 
and \emph{Canadian Journal of Mathematics, Canadian Mathematical Bulletin,  
Canadian Mathematical Communications} (Cambridge University Press). 
\end{IEEEbiographynophoto}

\vfill

\end{document}